\documentclass[fleqn,usenatbib]{mnras}

\usepackage[T1]{fontenc}
\usepackage{ae,aecompl}
\usepackage{totcount}
\newtotcounter{citnum} 
\def\oldbibitem{} \let\oldbibitem=\bibitem
\def\bibitem{\stepcounter{citnum}\oldbibitem}

\usepackage{graphicx}	
\usepackage{amsmath}	
\usepackage{amssymb}	
\usepackage{multicol}

\newcommand{\sgBe}{B[\,e\,] supergiant}
\newcommand{\sgBeshorthand}{B[\,e\,]SG}
\newcommand{\kms}{$\rm km\,s^{-1}$}
\newcommand{\phantomsph}{\texttt{PHANTOM}}

\newcommand{\lessthanapprox}{%
  \newcommand{\p}{%
    \setbox0=\vbox{\hbox{$<$}}%
    \ht0=0.6ex \box0 }%
  \newcommand{\s}{%
    \vbox{\hbox{$\sim$}}%
  }%
  \mathrel{\raisebox{0.7ex}{%
      \mbox{$\underset{\s}{\p}$}%
    }}%
}

\usepackage{newtxtext, newtxmath}

\title[GG Carinae's atomic circumbinary emission]{The circumbinary rings of GG Carinae: indications of disc eccentricity growth in the B[e] supergiant's atomic emission lines}

\author[A. J. D. Porter et al.]{
Augustus Porter$^{1}$\thanks{E-mail: augustusjdporter@gmail.com},
Katherine Blundell$^{1}$,
and Steven Lee$^{2,3}$
\\
$^{1}$Department of Physics, University of Oxford, Keble Road, Oxford, OX1 3RH, United Kingdom\\
$^{2}$Anglo-Australian Telescope, Coonabarabran NSW 2357, Australia\\
$^3$Research School of Astronomy and Astrophysics, Australian National University, Canberra, ACT 2611
}

\date{Accepted 2021 October 18. Received 2021 October 17; in original form 2021 February 02}
\pubyear{2020}

\hypersetup{final}
\begin{document}
\label{firstpage}
\pagerange{\pageref{firstpage}--\pageref{lastpage}}
\maketitle

\begin{abstract}
B[e] supergiants have unusual circumstellar environments which may include thin, concentric rings displaying atomic line emission. GG Carinae is a B[e] supergiant binary which exhibits such a geometry in its circumbinary environment. We study atomic emission lines arising from GG Carinae's circumbinary disc in FEROS spectra collected between 1998 and 2015. We find that semi-forbidden Fe\,II] and permitted Ca\,II emission are formed in the same thin circumbinary ring previously reported to have forbidden [O\,I] and [Ca\,II] emission. We find that there are two circumbinary rings orbiting with projected velocities of $84.6\pm1.0$\,\kms\ and $27.3\pm0.6$\,\kms. Deprojecting these velocities from the line-of-sight, and using updated binary masses presented by \cite{Porter2021GGPhotometry}, we find that the radii of the circumbinary rings are $2.8^{+0.9}_{-1.1}$\,AU and $27^{+9}_{-10}$\,AU for the inner ring and outer ring respectively. We find evidence of subtle dynamical change in the inner circumbinary ring over the 17 years spanned by the data, manifesting in variability in the ratio of the intensity of the blueshifted peak to the redshifted peak of its emission lines and the central velocity becoming more blueshifted. We perform smoothed-particle hydrodynamic simulations of the system which suggest that these observed changes are consistent with pumping of the eccentricity of a radially thin circumbinary ring by the inner binary. We find a systemic velocity of the GG Carinae system of $-23.2 \pm 0.4$\,\kms. 
\end{abstract}

\begin{keywords}
stars: binaries -- stars: emission-line, Be -- stars: supergiants -- stars: individual: GG Car
\end{keywords}



\section{Introduction}
\sgBe s (\sgBeshorthand s) are a class of rare stars which are  not predicted by stellar evolution theories. They are characterised by high luminosity and mass, hybrid spectra with infrared excess, strong emission in the Hydrogen Balmer series, forbidden and low-ionisation emission lines, and broad absorption in the ultraviolet for highly ionised species. These all indicate a complex circumstellar environment: the classical picture consists of a fast, hot, and ionised polar wind, and a slow, cool equatorial outflow \citep{Zickgraf1985, Zickgraf1986, Lamers1998, Kraus2019ASupergiants}. The thick, opaque circumstellar environments of \sgBeshorthand s generally preclude the direct observation of the stellar surfaces, making exact measurements of the stellar parameters difficult \citep{Kraus2009}. However, for the few examples for which it has been possible, they have been found to be rapid rotators \citep{Gummersbach1995BeClouds, Zickgraf1999TheStarscorrected, Zickgraf2006BeCloudscorrected, Kraus2008}; however, \cite{Kraus2016} have cast doubt on the accuracy of these rotational velocity measurements. The formation channels and the origin of the B[\,e\,] phenomenon are unclear, with some studies ascribing the phenomena to binarity \citep{Podsiadlowski2006corrected, Miroshnichenko2007, Wang2012AMBER/VLTI300} and others to non-radial pulsations \citep{Kraus2016}. \\

The classical \sgBeshorthand\ picture of an outflowing equatorial wind has been challenged in recent observations of circumstellar discs in Keplerian rotation around the stars, which are detached from the stellar surfaces \citep{DeWit2014DustyObservations}. CO and SiO bandhead emission has been observed in \sgBeshorthand s, indicating molecular emission in their cool outer discs \citep{Kraus2009, Kraus2013, Oksala2013ProbingTransition, Kraus2015}. Spectroscopy of forbidden emission lines has puzzlingly shown that \sgBeshorthand s are often surrounded by discrete, concentric, sometimes inhomogeneous, radially thin rings which emit the forbidden emission \citep{Kraus2010b, Wheelwright2012VLTI/AMBER327083, Aret2012, Kraus2016a, Torres2018a, Maravelias2018}. These emission lines are characterised by double-peaked Keplerian disc profiles (or partial disc profiles for inhomogeneous rings), often superposed upon each other, and each showing remarkably little radial extent. The thin radial extent can be observed by the disc profiles of the emission lines being acceptably modelled by annuli, and the disc profiles having sharp peaks in intensity and steep edges. These detached, inhomogeneous rings are inconsistent with the discs in \sgBeshorthand s being continual outflows from the stellar equators. Numerous methods have been invoked to explain these unusual distributions, such as: asymmetrical mass-loss due to rotation \citep{Zickgraf1985}; non-radial pulsations leading to mass-loss \citep{Kraus2016}; ``shepherd moon''-like massive bodies leading to gaps and rings in the circumstellar disc \citep{Kraus2016}; and binarity. \cite{Krauscorrected} reviews the different proposed formation mechanisms. \\

GG Carinae (GG Car) is a Galactic, confirmed binary with a \sgBeshorthand\ as its primary. Its well known $\sim$31-day orbital period has long been observed photometrically and spectroscopically \citep{Kruytbosch1930Variability2855, Greenstein1938FourCarinae, Hernandez1981, Gosset1984, Gosset1985, VanLeeuwen1998HipparcosVariables, Marchiano2012}. \cite{Porter2021GGPhotometry} used the spectroscopic variation of the emission lines from the wind to determine an accurate orbital solution of the binary and constrain the mass of the secondary, which is of unknown class. Table \ref{tab:ggcar_parameters} lists the known stellar parameters of GG Car and the binary separation and eccentricity. \cite{Porter2021GGBinary} investigated the variability of the system over short timescales, and found a 1.583-day periodicity present in both photometry and spectroscopy whose amplitude is modulated by the orbital phase of the binary.\\

\begin{table}
\centering          
\begin{tabular}{ l l}
\hline \hline
\\
   $d$ & $3.4^{+0.7}_{-0.5}$\,kpc\\
   $M_{\rm pr}$  & $24\pm4\,M_\odot$ \\
   $M_{\rm sec}$ & $7.2^{+3.0}_{-1.3}\,M_\odot$ \\
   
   $T_{\rm eff}$ & $23\,000 \pm 2\,000    $\,K \\
   $L_{\rm pr}$ & $1.8^{+1.0}_{-0.7}\times10^5\,L_\odot$ \\
   $R_{\rm pr}$ & $27^{+9}_{-7}\,R_\odot$ \\
   $a_{\rm bin}$ & $0.61\pm0.03$\,AU \\
   $e_{\rm bin}$ & $0.50^{+0.03}_{-0.03}$ \\
   $\omega_{\rm bin}$ & $339.87^{+3.10}_{-3.06}$\,$^\circ$ \\
\end{tabular}
\caption{Stellar and binary parameters of GG Car. \textit{Gaia} Data Release 2 distance, $d$, and stellar parameters of the primary in GG Car, where $M_{\rm pr}$ is the mass of the primary, $M_{\rm sec}$ is the mass of the secondary, $T_{\rm eff}$ is the effective temperature of the primary, $L_{\rm pr}$ is the luminosity of the primary, $R_{\rm pr}$ is the radius of the primary, $a_{\rm bin}$ is the binary semi-major axis, $e_{\rm bin}$ the binary orbital eccentricity, and $\omega_{\rm bin}$ is the binary's argument of periastron. All values taken from \protect\cite{Porter2021GGPhotometry} except $T_{\rm eff}$, which is taken from \protect\cite{Marchiano2012}.} 
\label{tab:ggcar_parameters}      
\end{table}

As is typical for \sgBeshorthand s, GG Car exhibits a complex circumstellar environment. It has been shown to have a dusty envelope \citep{Allen1973NearObjects, Marchiano2012}, Fe\,II and [Fe\,II] orbiting in a ring \citep{Swings1974Similarities94878., Allen1976TheExcesses} and CO emission arising from a circumbinary disc \citep{Kraus2013}. Forbidden [O\,I] and [Ca\,II] emission have been reported from at least two concentric circumbinary rings \citep{Maravelias2018}. \cite{Pereyra2009} found that the H-alpha emission is polarised, indicating that the emission originates at least in part from an optically-thin rotating disc. No spectral signatures from the primary or secondary's photospheres are directly observable in the visible range and no SiO emission has yet been reported in the system. Studies of the system's circumbinary disc show that it has an inclination of $\sim$60$^\circ$ \citep{BorgesFernandes2010THEEYES, Kraus2013}. \cite{Kraus2013} argues against the circumbinary disc being formed by Roche Lobe overflow, instead arguing that it is formed of material that was decreted from the primary while it was in a former evolutionary phase of a classical Be progenitor.\\

In the present study, we analyse GG Car's circumbinary atomic emission lines in high resolution FEROS spectra in the context of the updated orbital parameters and stellar masses reported in \cite{Porter2021GGPhotometry}. Section \ref{sec:observations} describes the FEROS observations; Section \ref{sec:circumbinary_lines} presents the detected circumbinary atomic emission lines; Section \ref{sec:model} describes the thin-ring model that is used to fit the circumbinary emission lines; Section \ref{sec:results} presents the results of the fitting; Section \ref{sec:discussion} discusses these results in the context of the new orbital solution of \cite{Porter2021GGPhotometry} and presents the results of smoothed-particle hydrodynamic simulations which can recreate the observed variability of the emission lines; and Section \ref{sec:conclusion} presents our conclusions. \\

\section{Spectroscopic observations}
\label{sec:observations}
The Fiber-fed Extended Range Optical Spectrograph (FEROS, \citealt{Kaufer1999CommissioningLa-Silla.}) is a high-resolution echelle spectrograph, located at the European Southern Observatory (ESO) at La Silla, Chile. Pre-October 2002, the spectrograph was used with the ESO 1.52m telescope; since then it has been used with the 2.2m MPG/ESO telescope. FEROS has a wavelength range of 3600\,-\,9200\,\AA\ and $R$\,$\sim$\,$48\,000$. The FEROS spectra used in this study have been reduced by ESO using the standard FEROS reduction pipeline. We calculate the expected atmospheric transmission in each FEROS spectrum using \texttt{TAPAS} \citep{Bertaux2014TAPASAstronomy}, and divide these from the spectra. FEROS observed GG Car in 1998 (8 spectra), 2009 (1 spectrum), 2011 (2 spectra), and 2015 (12 spectra). Any spectra taken on the same night are summed in order to increase signal to noise. Table \ref{tab:feros_observations} in the appendix lists the observation times of the FEROS spectra studied. Our preliminary studies of the circumbinary matter around GG Carina used the Global Jet Watch spectra \citep{Porter2021GGPhotometry, Porter2021GGBinary} but this paper exclusively presents results from the higher spectral resolution data from FEROS though these are consistent with one another. \\

\section{Atomic circumbinary emission lines}
\label{sec:circumbinary_lines}

We find the steep-sided, double-peaked line profiles associated with thin rings in a number of emission lines. In this Section we describe each of these atomic species and their line profiles. \\

\begin{figure}
  \centering
    \includegraphics[width=0.5\textwidth]{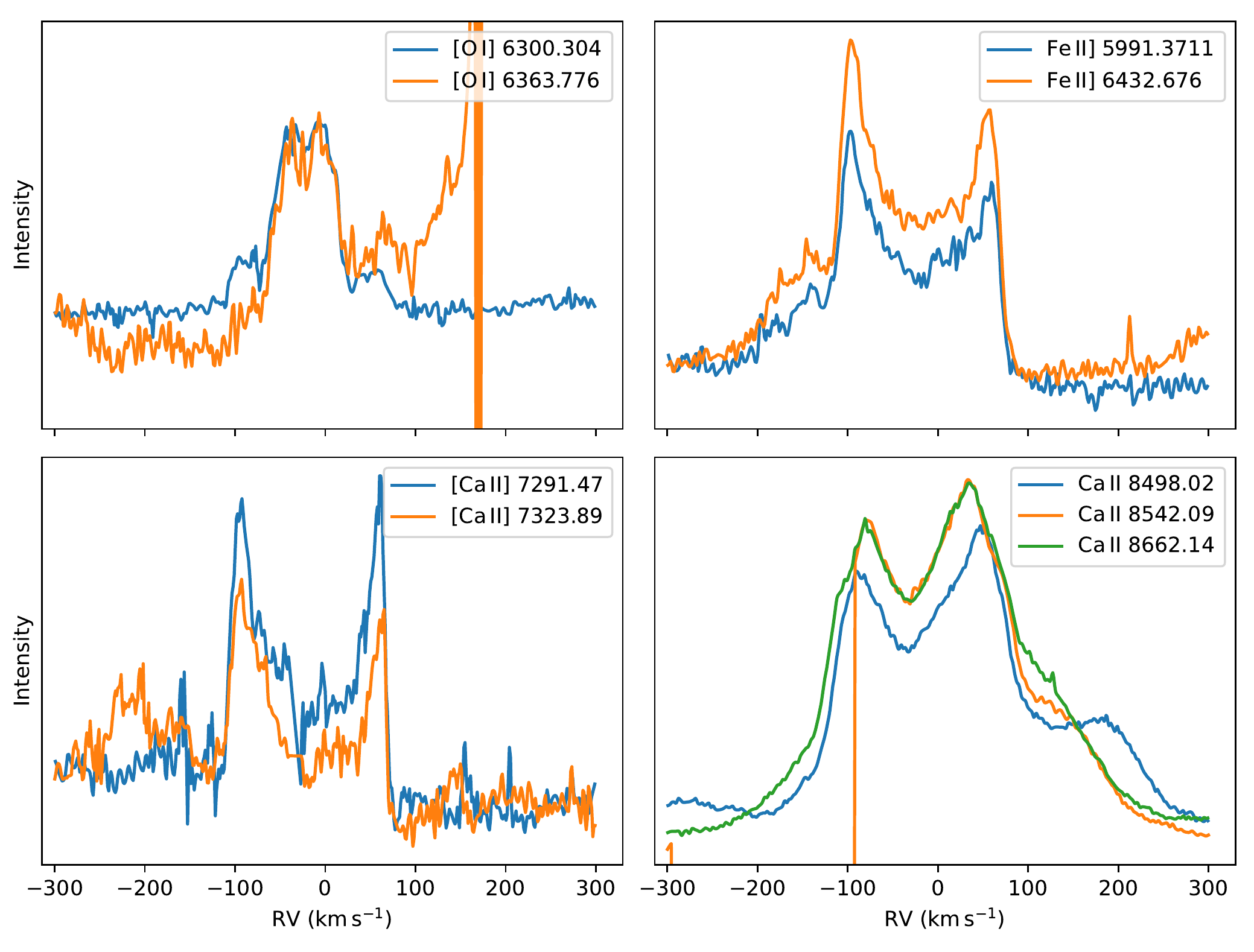}
    \caption{Atomic circumbinary ring profiles of GG Car, in radial velocity (RV) space, as observed by FEROS on 1998-12-06. Top left: [O\,I] emission. The [O\,I] 6364 line has Si\,II 6371 to its redward side. Bottom left: [Ca\,II] emission. Top right: Semi-forbidden Fe\,II] emission. Bottom right: Permitted Ca\,II emission. Every Ca\,II line in GG Car's spectrum is blended by a member of the H\,I Paschen series, which are to the redward side of Ca\,II. There is a data gap in the FEROS spectra for RV $<-95$\,\kms\ for the 8542\,\AA\ line.}
    \label{fig:ring_profiles}
\end{figure}

\subsection{[O\,I] emission}

We find [O\,I] 6300 and 6364\,\AA\ are present in the spectra of GG Car. Unlike some other \sgBeshorthand s, there is no [O\,I]\,5577\,\AA\ emission present. Figure \ref{fig:ring_profiles}, top left panel, displays the [O\,I] emission, as observed by FEROS. In the [O\,I] 6300 line, there are two distinguishable ring profiles superposed: one with a splitting $\sim$\,80\,\kms\ and the other with splitting $\sim$25\,\kms. This is indicative of two concentric rings giving rise to this emission. The [O\,I] 6364 line only exhibits a single ring profile, with a splitting similar to that of the lower-splitting 6300 profile. Since the 6364 line is slightly blended, and has low signal-to-noise, we only fit the [O\,I] 6300 line since it contains all of the relevant information from the 6364 line. \\

\subsection{[Ca\,II] emission}
In the FEROS spectra, [Ca\,II] 7291 and 7323\,\AA\ are present and display a single thin-ring profile. Figure \ref{fig:ring_profiles}, bottom left panel, displays the [Ca\,II] profiles observed by FEROS. Since both lines are of comparable quality and have similar profiles, we use both [Ca\,II] lines in our analysis, summing the lines in velocity space to improve signal-to-noise. \\

\subsection{Fe\,II] emission}

We report semi-forbidden Fe\,II] emission lines which are formed in the higher-velocity circumbinary ring of GG Car. Fe\,II] 5991 and 6432\,\AA\ display clear double-peaked profiles consistent having arisen from rings. Other Fe\,II lines, such as Fe\,II 6416.9, also host some indication of double-peaked ring profiles; however, they are unacceptably blended with emission from the primary's wind and therefore they are not analysed in this study. \\

Figure \ref{fig:ring_profiles}, top right panel, displays the profiles of the circumbinary Fe\,II] emission, observed by FEROS. There is a small emission component to the blueward side of the disc profile which causes an asymmetry in the profile; this component is likely to arise in the wind of the \sgBeshorthand\ primary. In our analysis, we fit this component and remove its contribution to focus on the disc profile. As for the [Ca\,II] lines, we sum the Fe\,II] lines in velocity space in the upcoming analysis to maximise signal-to-noise. \\

\subsection{Permitted Ca\,II emission}
We also report the discovery of permitted Ca\,II emission originating in GG Car's circumbinary disc; these are Ca\,II 8498.02, 8542.09, and 8662.14\,\AA. Coincidentally, each of these Ca\,II lines is blended with a different member of the H\,I Paschen series on their redward sides which are bright and of complex profiles. The Paschen series are associated with H\,I transitions from the $n$th to the third energy level, where $n > 3$. The polluting members are the $n=16$ ($\lambda = 8502$\,\AA), $n=15$ ($\lambda = 8545$\,\AA), and $n=13$ ($\lambda = 8665$\,\AA) transitions.\\

Figure \ref{fig:ring_profiles}, bottom right panel, presents the permitted Ca\,II emission, clearly displaying double-peaked disc profiles with similar splitting to the [Ca\,II] emission. The asymmetry and redward shoulder in the profiles is caused by the H\,I Paschen emission lines, which are $\sim$3\,\AA\ to the red of the Ca\,II lines. There is a data gap in the FEROS spectra to the blueward side of the Ca\,II 8542\,\AA\ line; therefore, we do not study this line further in this analysis. \\

In the November 2015 spectra, the double peaked profiles of the Ca\,II lines are too heavily polluted by the Paschen lines to successfully analyse.\\

\section{Thin ring model}
\label{sec:model}

For the purpose of fitting the orbital motion of the circumbinary rings to the emission lines, we assume that the material constituting the rings are on circular, quasi-Keplerian orbits about the binary. While in Section \ref{sec:disc:changes_in_v0} we show that circular symmetry may in fact be broken in the later spectra, this assumption allows an expeditious method to extract physically relevant parameters from the data without introducing the danger of over-fitting the data. \\

The Keplerian orbital speed, $v_{\rm kep}$, of a particle orbiting enclosed mass $M$ at radius $R$ is given by
\begin{equation}
\label{eq:keplerian}
    v_{\rm kep} = \sqrt{\frac{G M}{R}},
\end{equation}

\noindent where $G$ is the gravitational constant. The radial velocity of this particle towards the observer, $v_r$, is
\begin{equation}
\label{eq:radial_velocity}
    v_r = v_{\rm kep} \sin i \sin \theta + v_0,
\end{equation}

\noindent where $\theta$ is the true anomaly of the particle in its orbit (with $0^\circ$ defined as conjunction of the particle in front of the centre of mass), $i$ is the inclination of the orbit, and $v_0$ is the systemic radial velocity of the system. We assume that the ring is infinitesimally thin in the radial direction, and we use the terminology ``thin ring'' to mean radially thin for the remainder of this study. For a thin, circumbinary ring model, there are an infinite number of particles in orbit about $M$ with orbital speed $v_{\rm kep}$ distributed uniformly in $\theta$ from $0^\circ \rightarrow 360^\circ$. We assume each particle emits an equal amount of radiation making up the emission line profile, and that none of the emission is occulted. Therefore, the intensity of an emission line as a function of radial velocity $I(v_r) \propto P(v_r) \propto d\theta / d v_r$, where $P(v_r)$ is the distribution of $v_r$ given the uniform distribution of $\theta$. Therefore, from Equation \ref{eq:radial_velocity}, $I(v_r)$ can be analytically shown to be

\begin{equation}
\label{eq:ring_profile}
  I(v_r) \propto \left\{
  \begin{array}{@{}ll@{}}
    \frac{1}{\sqrt{\displaystyle1 - \left(\frac{v_r - v_0}{v_{\rm kep} \sin i}\right)^2}}, & \text{if}\ |v_r - v_0| < v_{\rm kep} \sin i \\ \\
    0, & \text{otherwise}.
  \end{array}\right.
\end{equation}

Equation \ref{eq:ring_profile} assumes perfect spectral resolution and an infinitesimally thin Keplerian-orbiting ring with no turbulence or thermal broadening, none of which will be strictly true. In order to get a real line profile from the theoretical disc profile defined in Equation \ref{eq:ring_profile}, $I(v_r)$ needs to be convolved with a Gaussian of standard deviation $\sigma$; $\sigma$ encodes the resolution of the observing instrument, turbulent velocity in the ring, thermal velocities in the ring, and any finite radial width of the emission region. We construct the line profiles by numerically convolving Equation \ref{eq:ring_profile} with a Gaussian of width $\sigma$. \\

We fit ring profiles to the [O\,I], Fe\,II], [Ca\,II], and Ca\,II emission lines by using the Monte Carlo Markov Chain algorithm \texttt{emcee} \citep{Foreman-Mackey2012Emcee:Hammer} to determine the posterior probability distributions of the ring parameters. We fit for the physically relevant parameters $v_{\rm kep} \sin i$, $v_0$, and $\sigma$, along with the base continuum level, local continuum gradient, and peak intensity, which are physically insignificant parameters in this study and are therefore excluded from further analysis. The line intensities are a reflection of the continuum brightness which is variable at both the orbital period and 1.583-day period (Papers A and B), and therefore have little physical significance when considered on their own in GG Car. \\ 

\begin{figure}
  \centering
    \includegraphics[width=0.5\textwidth]{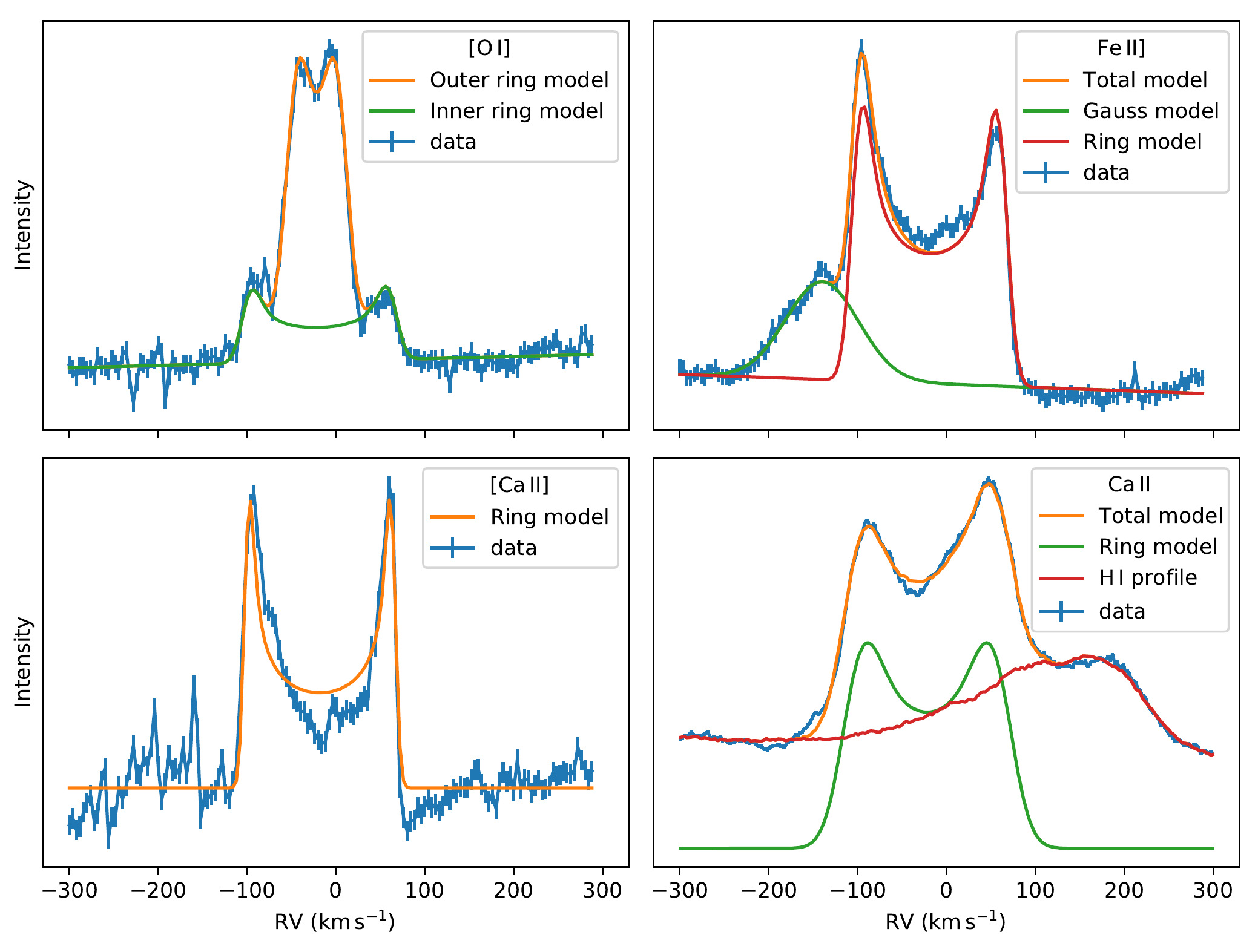}
    \caption{Example fits to the circumbinary ring lines. All profiles were observed by FEROS on 1998-12-07. Top left: Double circumbinary ring fit to the [O\,I] 6300 line. Top right: ring and Gaussian fit to the summed Fe\,II] 5991.4 and 6432.7\,\AA\ lines. Bottom left: single ring fit to the summed [Ca\,II] 7291.5 and 7323.9\,\AA\ lines. Bottom right: circumbinary ring and combined H\,I Paschen profile fit to the Ca\,II 8498.0\,\AA\ line.}
    \label{fig:ring_profiles_fits}
\end{figure}

As the [O\,I] 6300 line is emitted from two separate circumbinary rings, we fit this line as a superposition of two ring profiles, fitting for the parameters of both rings concurrently. Figure \ref{fig:ring_profiles_fits}, top left panel, displays an example of the [O\,I] 6300 line as observed by FEROS, and the double-ring profile fitted to the line. \\

The Fe\,II] 5991 and 6432 lines are mildly blended with another component likely to be formed in the wind of the primary. We model this polluting component as a Gaussian in the fitting routine, and the fitted Gaussian component is then subtracted from the data before further analysis is undertaken. Figure \ref{fig:ring_profiles_fits}, top right panel, displays our fit to the example FEROS spectrum's Fe\,II] lines. \\

As the [Ca\,II] lines are unpolluted by any components, they are fitted with a single ring profile. An example fit is shown in Figure \ref{fig:ring_profiles_fits}, bottom left panel. The Telluric correction routine does not fully remove the atmospheric absorption from the [Ca\,II] lines in the 1998-12-23 and 1998-12-24 spectra (see Figure \ref{fig:feros_line_overlay}, top row, rightmost two panels). We are still able to extract acceptable fits from these lines which are consistent with the spectra from near-in-time spectra, therefore we retain these fits for further analysis.\\

The permitted Ca\,II lines are polluted by the H\,I Paschen series. To account for this, for each spectrum we build an average Paschen line profile in RV space using the unpolluted and strong Paschen lines in the FEROS wavelength range. These are the $n = $ 12, 14, and 17 lines. We then fit the circumbinary ring profile with the Paschen profile at its appropriate RV offset, with the intensity of the Paschen profile as a free parameter. Figure \ref{fig:ring_profiles_fits}, bottom right panel, shows an example fit to the Ca\,II 8498.0\,\AA\ line. \\

\section{Results}
\label{sec:results}

\begin{figure}
  \centering
    \includegraphics[width=0.5\textwidth]{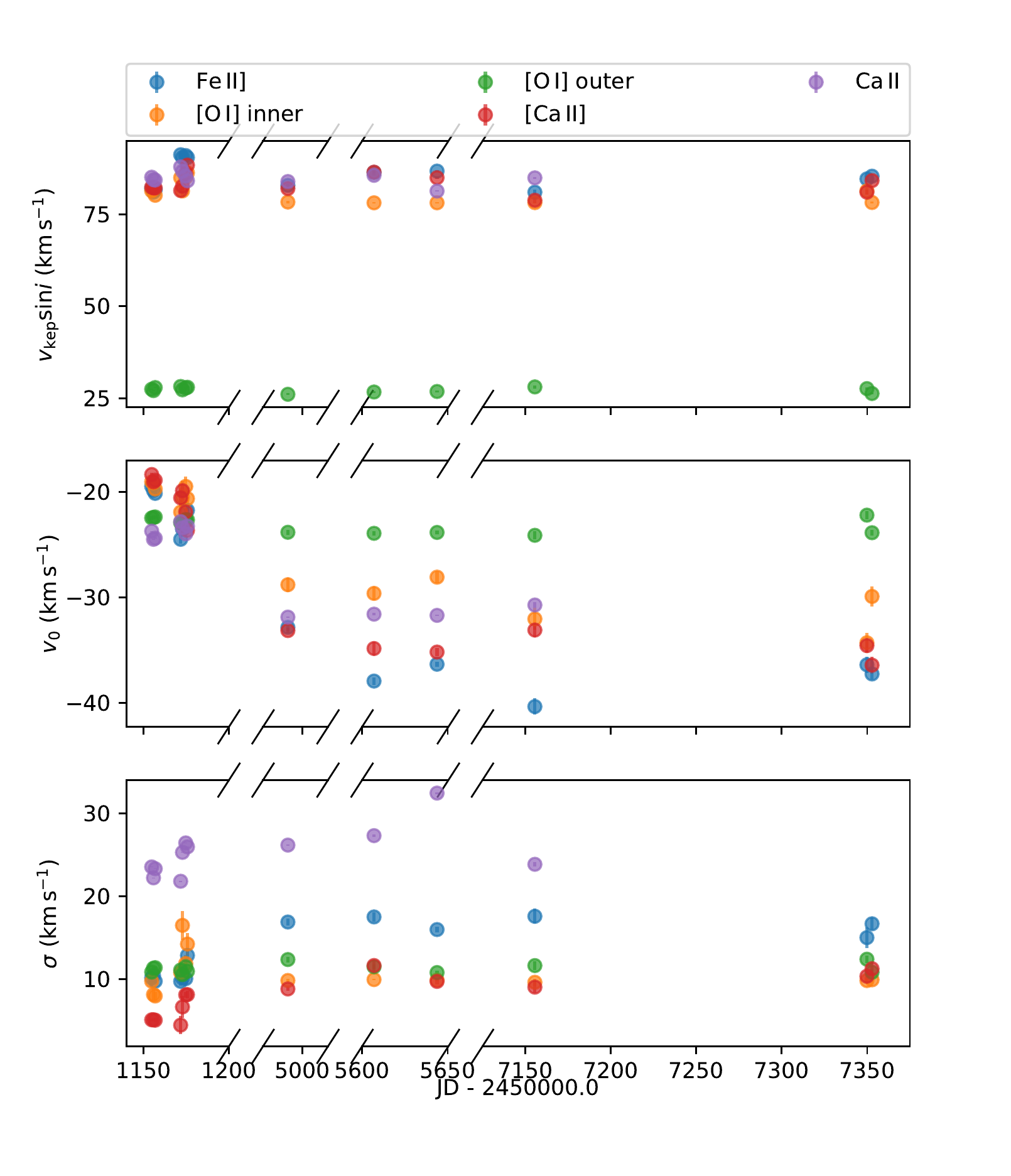}
    \caption{Fitted projected Keplerian velocity, $v_{\rm kep} \sin i$, central velocity, $v_0$, and convolution width, $\sigma$, against time for [O\,I] 6300, [Ca\,II], Ca\,II, and Fe\,II] lines in the FEROS spectra against time of observation. Note the breaks in the $x$ axis, which split the spectra by year of observation (1998, 2009, 2011 and 2015). }
    \label{fig:feros_fitted_ring_parameters}
\end{figure}

Figure \ref{fig:feros_fitted_ring_parameters} displays the fitted parameters of the FEROS emission lines against time of observation. Note that the $x$ axis skips periods of time without observations. The FEROS fits during 2015 are included, for comparison. There is little significant variability of the fitted ring parameters, \textcolor{black}{with the exception of the centroid velocity $v_0$ of the lines which have higher $v_{\rm kep} \sin i$ which tends to decrease over the FEROS observations}. The variability of the $v_0$ parameters of these lines are presented in more detail in Section \ref{sec:v0_variability}. The other fitted parameters show no significant changes over the FEROS dataset. The measured $\sigma$ of the Ca\,II lines is consistently larger than the other emission lines; this is discussed in Section \ref{sec:discussion}.\\

It is clear from Figure \ref{fig:feros_fitted_ring_parameters} that there are two families of $v_{\rm kep} \sin i$ values. This indicates that there are two circumbinary rings; an inner ring with higher $v_{\rm kep} \sin i$, which is manifested in all studied emission lines, and an outer ring with lower $v_{\rm kep} \sin i$, which is only manifested in [O\,I]. This is in agreement with \cite{Maravelias2018}. The measured $v_{\rm kep} \sin i$ values of the Fe\,II] and permitted Ca\,II emission are consistent with those of the inner ring of the [O\,I] and [Ca\,II] emission, and also the CO emission presented in \cite{Kraus2013}. This indicates that these line components originate in the same circumbinary region.\\

\begin{figure*}
  \centering
    \includegraphics[width=\textwidth]{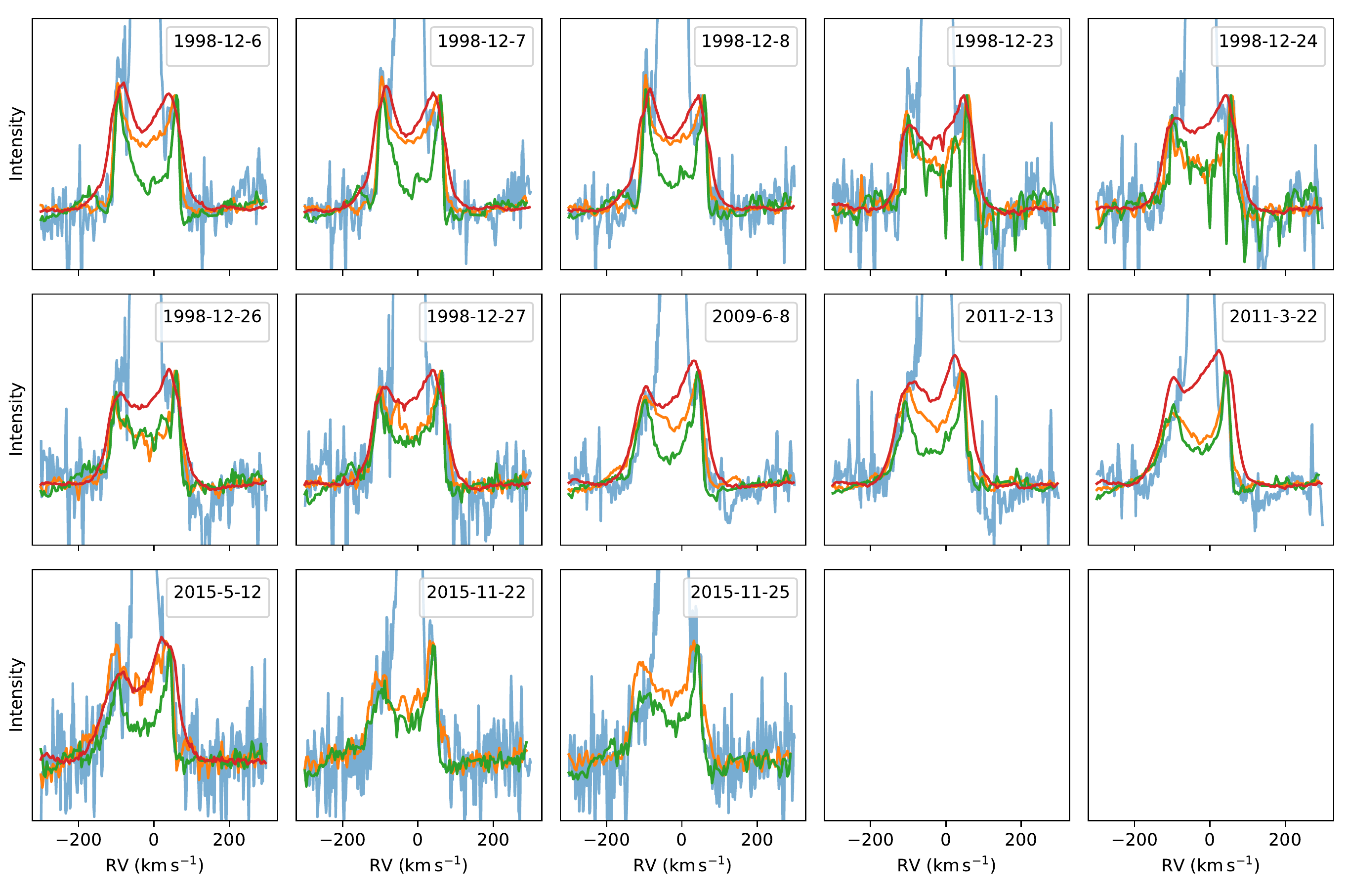}
    \caption{Overlay of the circumbinary emission lines in each of the FEROS spectra. [O\,I]\,6300\,\AA\ is plotted in blue, Fe\,II] in orange, [Ca\,II] in green, and Ca\,II in red. The Fe\,II] lines have had the polluting Gaussians subtracted, and the Ca\,II have had the polluting H\,I Paschen profiles subtracted. The profiles have been normalised by the intensity of the blueshifted peak of the profiles, and are zoomed to focus on the inner circumbinary ring profiles.}
    \label{fig:feros_line_overlay}
\end{figure*}

Figure \ref{fig:feros_line_overlay} displays the circumbinary emission profiles for each of the studied lines in each FEROS spectrum. The profiles have been processed such that the polluting Gaussian has been subtracted from the Fe\,II] emission lines, and the H\,I Paschen profiles have been subtracted from the Ca\,II lines. The emission line profiles have all been normalised by the red peak of the higher $v_{\rm kep}$ ring profile in Figure \ref{fig:feros_line_overlay}. The emission line profiles are remarkably similar and display the same evolution for all of the emission lines in the FEROS spectra, further indicating that these lines form in the same thin circumbinary region. The emission lines have slight evolution of $V$/$R$ over the FEROS observations. \\

\subsection{Average of fitted parameters}
Table \ref{tab:ring_parameters} lists the average values of the fitted ring parameters of all the line species for the inner and outer circumbinary rings. In Table \ref{tab:ring_parameters} we quote the average values of the fit parameters of the inner ring. The average $v_{\rm kep} \sin i$ is calculated using all of the ring profiles arising in the inner ring. The average $v_0$ also takes into account the $v_0$ of the [O\,I] from the outer ring, such that it indicates the systemic velocity of the system using all circumbinary lines. The average $\sigma$ only takes into account the inner ring's [O\,I], [Ca\,II], and Fe\,II], since the $\sigma$ of the Ca\,II is considerably larger. Therefore the $\sigma$ values of the [O\,I], [Ca\,II] and Fe\,II] lines observed by FEROS give the most representative $\sigma$ of the inner ring.\\

From these circumbinary lines we determine that the systemic velocity of the system is $-23.2\pm0.4$\,\kms. \\

\begin{table} 
\centering          
\begin{tabular}{l l l l}
\hline \hline
\\
    Line & $v_{\rm kep} \sin i$ (\kms) & $v_0$ (\kms) & $\sigma$ (\kms)\\ 
    \hline
    \textbf{Outer ring} & & & \\
    {[}O$\,$I{]} & $27.3 \pm 0.6 $ & $ -23.0 \pm 0.7 $ & $ 11.2 \pm 0.4 $ \\
    \\
    \textbf{Inner ring}  & & & \\
    Fe$\,$II{]} & $84.6 \pm 3.5 $ & $ -25.0 \pm 7.2 $ & $ 11.7 \pm 2.8 $ \\
    {[}O$\,$I{]} & $82.6 \pm 2.6 $ & $ -23.0 \pm 5.0 $ & $ 9.9 \pm 0.4 $ \\
    {[}Ca$\,$II{]} & $82.9 \pm 1.9 $ & $ -22.7 \pm 6.3 $ & $ 6.6 \pm 2.0 $ \\
    Ca$\,$II & $85.1 \pm 0.8 $ & $ -24.0 \pm 1.4 $ & $ 24.2 \pm 2.1 $ \\
    \textbf{Average} & $84.6\pm1.0$ & $-23.2\pm0.4$* & $9.8\pm0.7$ **  \\
\end{tabular}
\caption{Average ring profile fit parameters to the different line species. $v_{\rm kep} \sin i$ is the fitted Keplerian velocity which has not yet been deprojected from the line of sight, $v_0$ the centre velocity, and $\sigma$ the convolution Gaussian width. The values are the means of all the fit parameters in the different spectral observations weighted by the inverse fit uncertainties, and the errors are the standard deviations of the fitted parameters weighted by the inverse fit uncertainties. The average values of the inner ring's line species are shown. *The $v_0$ average takes into account the $v_0$ of the outer ring [O\,I]. ** The $\sigma$ average of the inner ring does not take into account Ca\,II.}
\label{tab:ring_parameters}      
\end{table}

\subsection{Central velocity changes of the inner ring}
\label{sec:v0_variability}
\begin{figure}
  \centering
    \includegraphics[width=0.5\textwidth]{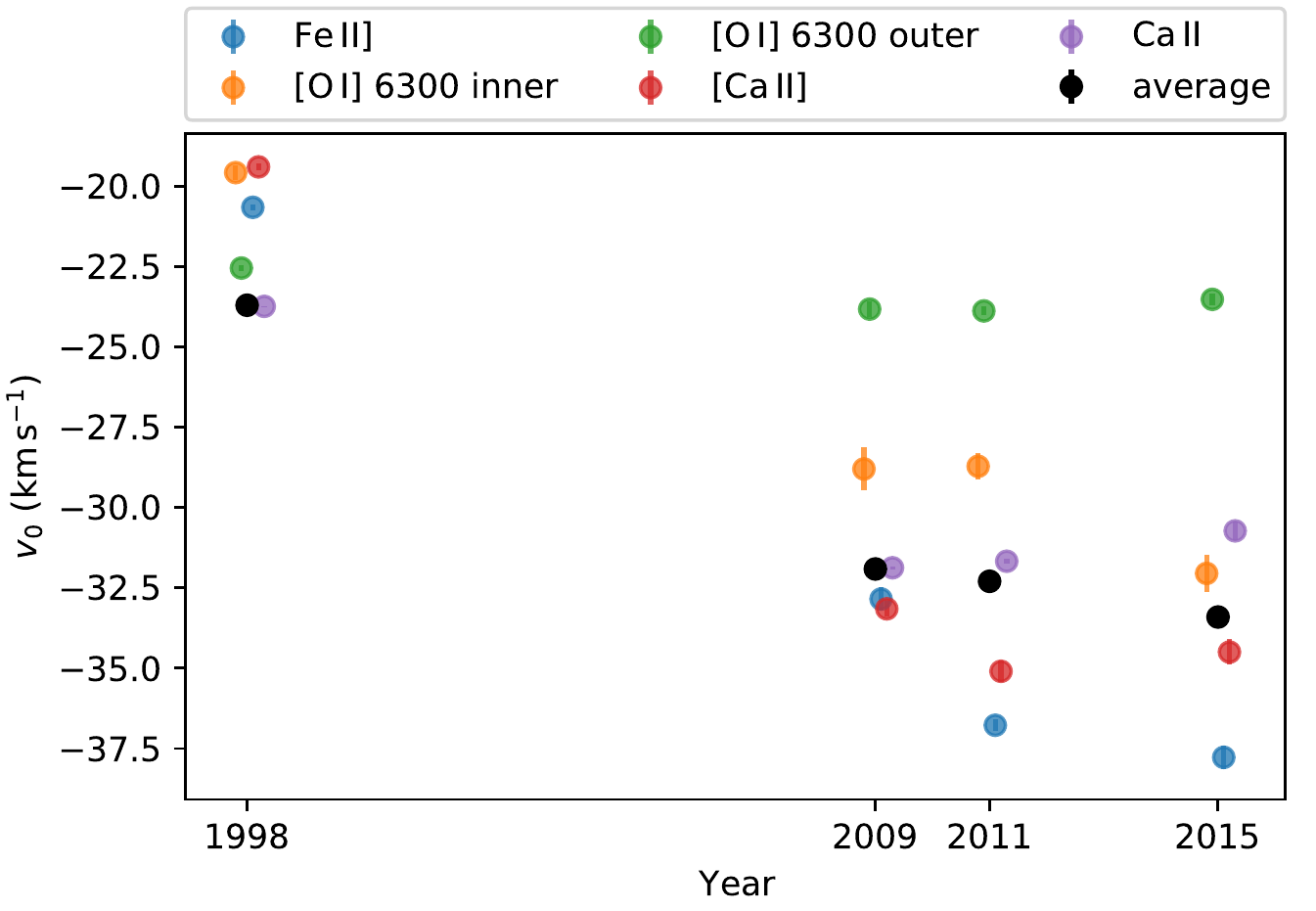}
    \caption{The changes in $v_0$ of the circumbinary rings as observed by FEROS. The fitted $v_0$ for each line species is averaged in the 1998, 2009, 2011, and 2015 spectra separately. Black points denote the average $v_0$ for the lines originating in the inner circumbinary ring.}
    \label{fig:v0_1998_2015}
\end{figure}

The only fitted parameter to significantly change during observations is the central velocity, $v_0$, of the inner ring's emission lines in the FEROS spectra. Figure \ref{fig:v0_1998_2015} displays the average $v_0$ of each line from each year of observation separately, and also displays the average $v_0$ for all of the lines consisting of the inner circumbinary ring as a black point. The average $v_0$ for the inner ring in 1998 is $-21.2\pm0.1$\,\kms, whereas it is $-32.0\pm0.1$\,\kms\ and $-34.5\pm0.1$\,\kms\ in 2009 and 2011, and $-33.4\pm0.2$\,\kms\ in 2015. The $v_0$ of the outer [O\,I] ring, on the other hand, is not significantly different throughout the observations, with a value of $-22.6\pm0.1$\,\kms\ in 1998, $-23.8\pm0.3$\,\kms\ in 2009, $-23.9\pm0.14$\,\kms\ in 2011, and $-23.5\pm0.2$\,\kms\ in 2015. The variability of the inner rings' $v_0$, and constancy of the outer ring's $v_0$, is reflected in the standard deviations of their constituent emission lines' $v_0$ values in Table \ref{tab:ring_parameters}; this leads to the outer [O\,I] ring making the strongest contribution to the average $v_0$ quoted in the bottom line of Table \ref{tab:ring_parameters}. Figure \ref{fig:vkep_1998_2015}, in the appendix, shows that the $v_{\rm kep} \sin i$ determined from the inner ring between 1998 and 2015 remains roughly constant. \\

\begin{figure}
  \centering
    \includegraphics[width=0.5\textwidth]{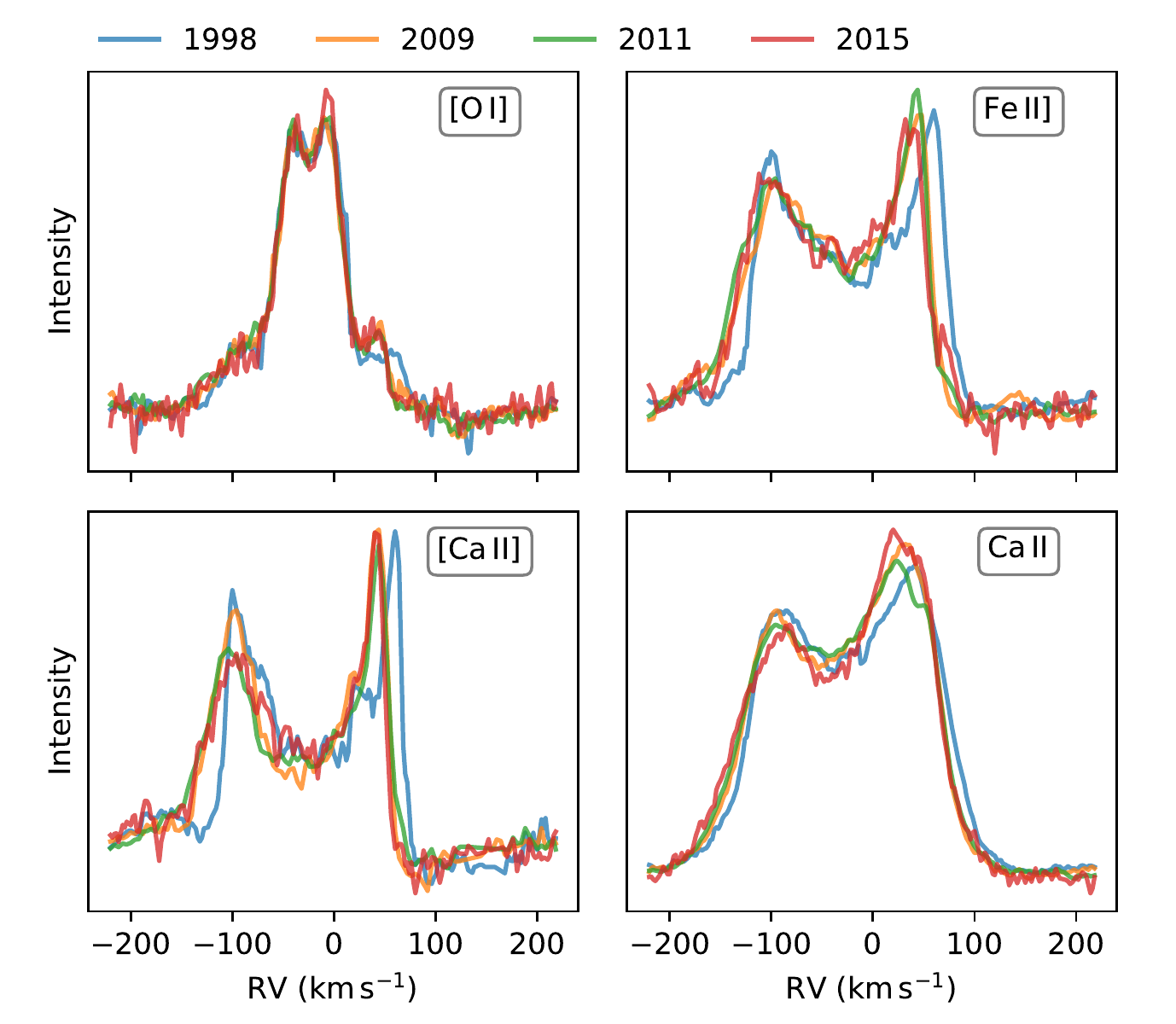}
    \caption{Summed circumbinary line profiles from FEROS, separating the spectra by year of observation. The Fe\,II] lines have had the polluting Gaussian components subtracted and the Ca\,II have had the polluting Paschen H\,I profiles subtracted before summation.}
    \label{fig:profiles_1998_2015}
\end{figure}

This pattern of the inner ring becoming more blueshifted over the FEROS observations is clear when circumbinary ring profiles of each line are averaged over all observations in a year. Figure \ref{fig:profiles_1998_2015} shows the circumbinary emission line profiles as observed by FEROS, with all spectra from the same year summed. In the $\sim$17 year time span of these observations, the higher velocity ring profile has become more blueshifted in each emission line. Correspondingly, there is some development in the shapes of the profiles, with the red peaks having higher intensity than the blue peak causing a decreasing $V$/$R$ ratio. The bluemost edge of the profiles also become less steep over the observations. For the lower velocity ring of [O\,I] 6300, however, the profile and central radial velocity are unchanged between 1998 and 2015. We verify that the $v_0$ shift of the inner ring is real, and not due to inconsistent wavelength calibrations, by removing the heliocentric velocity correction of the FEROS spectra and confirming that the telluric absorption features are at consistent wavelength. The relative drift between the central velocities of the [O\,I] and [Ca\,II] lines was also noted in \cite{Maravelias2018}. \\

\subsection{$V$/$R$ variations}

\begin{figure}
  \centering
    \includegraphics[width=0.5\textwidth]{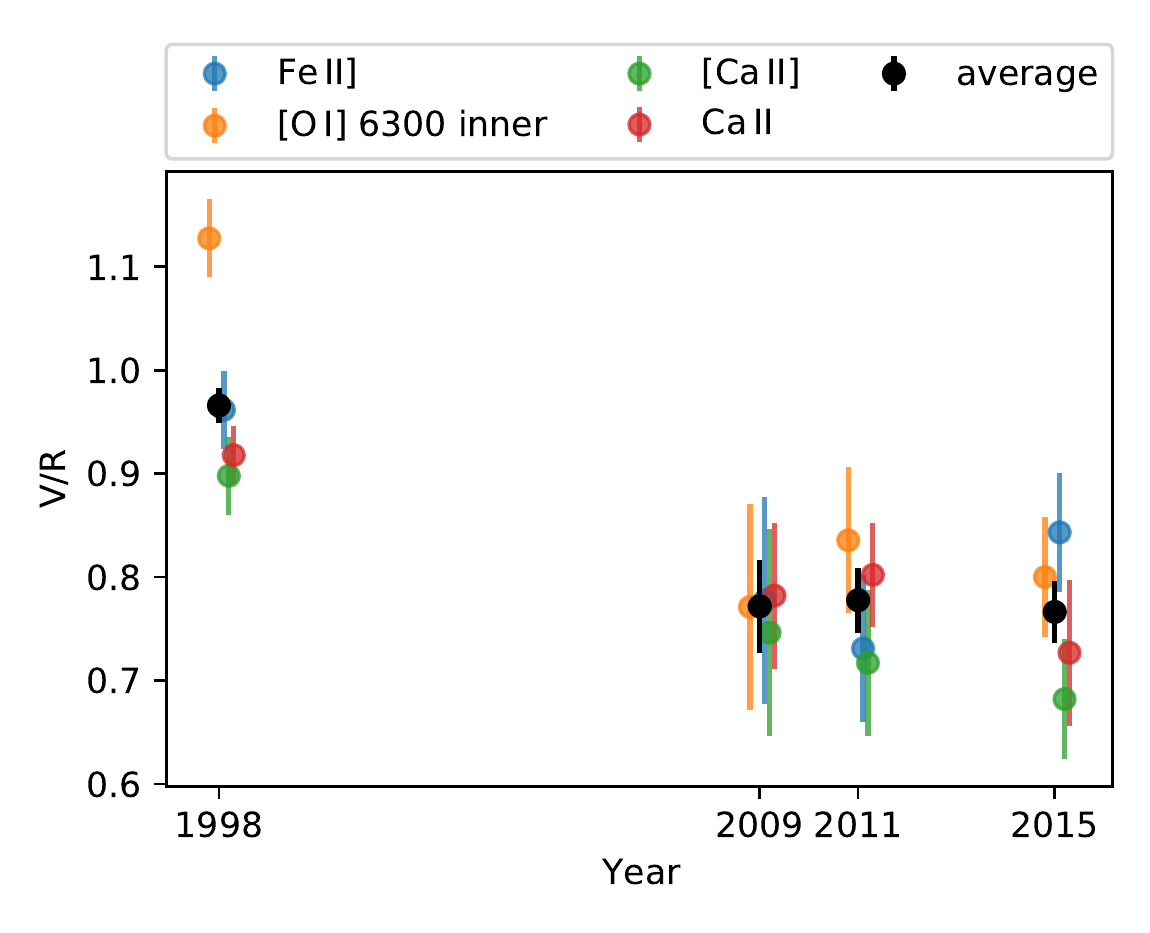}
    \caption{Same as Figure \ref{fig:v0_1998_2015}, except showing the evolution of the $V$/$R$ of the double-peaked emission lines arising from the inner circumbinary ring, as observed by FEROS, in each year of observation. The black points denote the mean $V$/$R$ values of all lines in a given year.}
    \label{fig:v_o_r_1998_2015}
\end{figure}

Figure \ref{fig:v_o_r_1998_2015} displays the $V$/$R$ peak ratios for the emission lines that originate in the inner circumbinary ring, observed by FEROS, against observation year. The Fe\,II] and Ca\,II lines had the polluting Gaussian component and H\,I Paschen profiles, respectively, subtracted before calculating the $V$/$R$ ratio. There is some slight variation in the $V$/$R$ parameter for the lines with the values dropping over time for each line, indicating that the red peaks are getting stronger compared to the blue peaks of the profiles. The varying $V$/$R$ of the profiles over time may also be seen by eye in Figure \ref{fig:profiles_1998_2015}.\\

\section{Discussion}
\label{sec:discussion}
For the higher $v_{\rm kep}\sin i$ lines listed in Table \ref{tab:ring_parameters}, their $v_{\rm kep}\sin i$ values agree with each other very well, therefore we can infer that these lines arise in the same radially thin circumbinary region about the binary. This is supported by Figure \ref{fig:feros_line_overlay}, which shows that all of the emission lines display remarkably similar circumbinary ring profiles throughout the FEROS observations. Therefore, in the upcoming analysis, we will consider the average values of $v_{\rm kep}\sin i$ for all lines in our calculations. \\

\subsection{Radii of the circumbinary rings}
\label{sec:circumbinary_disc_consequences}

In this analysis, we use the circumbinary disc inclination for the molecular ring of $60\pm20^\circ$, which is informed by the disc studies of \cite{BorgesFernandes2010THEEYES} and \cite{Kraus2013}.\\

\begin{table} 
\centering          
\begin{tabular}{l l l l}
\hline \hline
\\
    Ring & $v_{\rm kep} \sin i$ (\kms) & $v_{\rm kep}$ (\kms) & $R$ (AU)\\ 
    \hline
    Outer ring & $27.3 \pm 0.6$ & $32^{+11}_{-4}$ & $27^{+9}_{-10}$ \\
    \\
    Inner ring  & $84.6\pm1.0$ & $98^{+33}_{-11}$ & $2.8^{+0.9}_{-1.1}$ \\
\end{tabular}
\caption{Measured projected orbital speeds ($v_{\rm kep} \sin i$), deprojected orbital speeds ($v_{\rm kep}$), and circumbinary ring radius ($R$). $v_{\rm kep} \sin i$ are the same values as in Table \ref{tab:ring_parameters}.}
\label{tab:ring_orbital_speed_radius}      
\end{table}

Table \ref{tab:ring_orbital_speed_radius} lists the calculated deprojected orbital speeds and orbital radii of the circumbinary rings. They were calculated using Monte Carlo sampling, with 1\,000\,000 random Gaussian-distributed samples of $M_{\rm pr}$, $M_{\rm sec}$, $i$, and $v_{\rm kep} \sin i$ for both the inner and outer rings. $v_{\rm kep} \sin i$ are the average fitted Keplerian velocities to the ring profiles, and the values are taken from Table \ref{tab:ring_parameters}. For each sample, $v_{\rm kep}$ is calculated for each ring, and then the corresponding ring radius is calculated using the total enclosed mass of the binary and Equation \ref{eq:keplerian}. The median value of the parameter samples is taken as the best estimate value, and the difference of the median with the 84th and 16th percentiles is taken as the upper and lower uncertainties respectively. We find that the closer ring has a radius of $R = 2.8^{+0.9}_{-1.1}$\,AU from the binary, and the outer ring has $R = 27^{+9}_{-10}$\,AU from the binary. As expected, the rings orbit slightly closer into the binary than as reported by \cite{Maravelias2018} due to the lower masses of the binary components reported in \cite{Porter2021GGPhotometry}. These correspond to $\sim$\,4.7 and 45 semi-major axes of the binary, respectively. \cite{Marchiano2012} finds that the dust envelope begins at $\sim230 \, R_{\rm pr}$. Using $R_{\rm pr} = 27^{+9}_{-7}\,R_\odot$ from \cite{Porter2021GGPhotometry}, this corresponds to the dust envelope radius of $29^{+10}_{-7}$\,AU. This places the outer O\,I ring in a similar region to the inner edge of the system's dust envelope.\\

The $\sigma$ parameters of the forbidden emission lines observed by FEROS shown in Table \ref{tab:ring_parameters} indicate that the circumbinary emission originates in radially thin rings. The velocity resolution of FEROS is $\sigma_{\rm res} = c / R = 6.25$\,\kms\ and, with turbulent and thermal velocities, these can account for the majority of the measured $\sigma$s of these lines. The permitted Ca\,II emission FEROS has a much larger $\sigma$ than the forbidden emission lines. This may be due to the permitted emission originating in a more radially extended region than the forbidden emission, which requires the correct conditions of density and temperature in order to be emitted. We can very roughly make an estimate of the difference in radial width of the Ca\,II and forbidden emitting regions by comparing the $\sigma$ values. $\sigma$ is composed of multiple constituents, such that
\begin{equation}
    \sigma^2 = \sigma_{\rm res}^2 + \sigma_{\rm therm}^2 + \sigma_{\rm turb}^2 + \sigma_{\rm size}^2,
\end{equation}

\noindent where $\sigma_{\rm res}$ is the spectral resolution of the observing instrument, $\sigma_{\rm therm}$ is the thermal broadening, $\sigma_{\rm turb}$ is broadening due to turbulence, and $\sigma_{\rm size}$ is broadening due to a finite radial width of the circumbinary rings. Assuming that $\sigma_{\rm therm} \approx \sigma_{\rm turb} \approx 1$\,\kms, we can make an estimate for $\sigma_{\rm size}$. If we then approximate $\sigma_{\rm size}$ as of the order of the orbital velocity difference between the inner and outer edge of a circumbinary ring, we can show that

\begin{equation}
\left(\frac{\sigma_{\rm size}}{\sin i}\right)^2 \approx \left(v_{\rm kep}^{\rm in} - v_{\rm kep}^{\rm out}\right)^2 = GM\,\frac{R_{\rm out} - R_{\rm in}}{R_{\rm in}R_{\rm out}} \approx GM\,\frac{\Delta R}{R^2},
\end{equation}

\noindent where $v_{\rm kep}^{\rm in}$ and $v_{\rm kep}^{\rm out}$ are Keplerian velocities of the inner and outer edges respectively, $R_{\rm in}$ and $R_{\rm out}$ are the radii of the inner and outer edges respectively, $\Delta R = R_{\rm out} - R_{\rm in}$, and $R$ is the representative radius taken from the double-peaked Keplerian ring profile. $\Delta R$ is therefore the representative width of the ring. Using the average $\sigma$ of the inner ring's forbidden emission (from the bottom line of Table \ref{tab:ring_parameters}) gives $\Delta R = 0.02\pm0.01$\,AU, whereas using the $\sigma$ of the permitted Ca\,II emission gives $\Delta R = 0.2\pm0.1$\,AU. It must be noted that, since it originates from permitted transitions, the Ca\,II emission may suffer some self-absorption and re-emission which may also be a contributing factor to its larger $\sigma$. We remark that this method only yields very rough estimates for the relative radial widths of the rings.\\

\subsection{Limits on precession of the circumbinary ring's inclination}
Test particles' orbits in circumbinary trajectories are expected to precess about the inner binary in inclination and longitude of ascending node, dependent on the mass ratio, eccentricity, and separation of the binary \citep{Doolin2011}. The same precession occurs in hydrodynamic circumbinary disc simulations; however, the amplitude of the discs' precession becomes damped over time and the inclination and longitude of ascending node reach a steady state \citep{Martin2017, Martin2018, Martin2019PolarMass, Smallwood2019Alignment15D}. \\

For the thin circumbinary rings, precession in the inclination of their orbits to the line of sight would be observed as variations in the measured $v_{\rm kep} \sin i$ and the splitting of the peaks of the emission line profiles as the inclination and longitude of ascending node of the ring varies around the binary. We do not witness any significant variations in the fitted $v_{\rm kep} \sin i$ over the FEROS dataset. Given $v_{\rm kep} \sin i = 84.6\pm1.0$\,\kms\ for the inner ring, we can limit the changes to the inclination of the inner circumbinary ring's orbit to the line of sight, $\Delta i$, to $\lessthanapprox 7^\circ$, assuming an inclination of $\sim$60$^\circ$ over the observational period. \\

\subsection{An eccentric ring?}
\label{sec:disc:changes_in_v0}
The changes in $v_0$ observed in the inner circumbinary ring, described in Section \ref{sec:v0_variability}, between 1998 and 2015 data are unusual, since a circumbinary disc is expected to have a constant systemic velocity equal to that of the centre of mass of the system. The ring cannot be circumstellar, since its $v_{\rm kep}$ value and the individual masses of the binary components imply a ring radius that is greater than the orbital separation of the binary, by Equation \ref{eq:keplerian}. Therefore, the $v_0$ change cannot be a reflection of the binary orbit. \\

Presuming that the circumbinary ring cannot have a true bulk velocity difference to the centre of mass of the system, and the constant $v_0$ of the outer ring observed in [O\,I] implies that centre of mass velocity of the system is unchanged, a change in $v_0$ implies that circular symmetry of the inner circumbinary ring has been broken over the course of the FEROS observations. An axisymmetric ring orbiting the centre of mass will have the same central velocity at the centre of mass, and will lead to a symmetric and constant profile, which is clearly not observed. \\

Two effects may lead to a loss of circular symmetry in the circumbinary rings: either overdensities are present in the ring, or the circumbinary ring itself is eccentric. With only a few epochs of observation of the FEROS spectra, it is difficult to determine exactly which process is occurring in the circumbinary ring. However, eccentricities are expected to be excited in circumbinary discs due to interaction with the binary \citep{Lubow2000Interactionsdisks}, and hydrodynamic simulations of circumbinary discs consistently show that this is the case, and furthermore that their eccentricities may even be caused to oscillate (e.g. \citealt{Papaloizou2001OrbitalInteraction, Pierens2013, Dunhill2015, Lines2015, Lines2016, Fleming2016, Thun2017}). The densities of these simulated eccentric discs are higher at apastron. Therefore, in GG Car, the varying $v_0$ and $V$/$R$ of the inner circumbinary ring lines may indicate that the binary has pumped the eccentricity of the ring.\\

\begin{figure}
  \centering
    \includegraphics[width=0.5\textwidth]{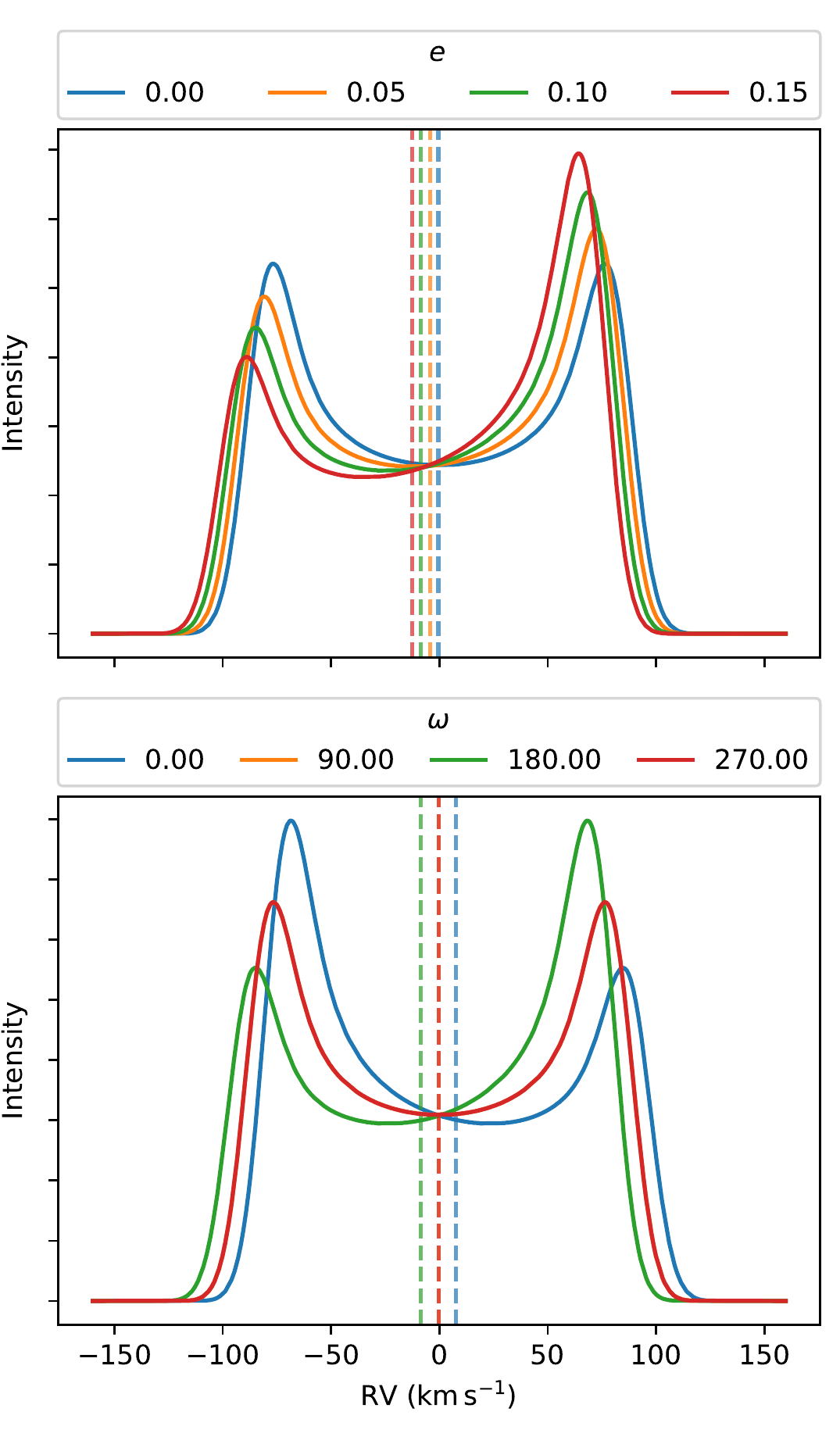}
    \caption{Effect of varying eccentricity, $e$ (top panel), and argument of periastron, $\omega$ (bottom panel), of a circumbinary ring on the observed emission line profile. The amplitude of the rings are kept constant at $84.6$\,\kms, and the profiles are convolved with a Gaussian of width 9.8\,\kms. $\omega$ in the top panel is kept constant at 180$^\circ$, and $e$ in the bottom panel is kept constant at 0.1. Vertical dashed lines with a matching colour of a model indicate the mid-point between the two peaks of the given model. Note that the line profiles for $\omega = 90^\circ$ and $\omega = 270^\circ$ are identical.}
    \label{fig:profile_change_eccentricities}
\end{figure}

Figure \ref{fig:profile_change_eccentricities}, top panel, displays a simple model of what changes in line profiles may be expected if eccentricities are pumped in a thin circumbinary ring. For a range of eccentricities, the orbital velocity amplitude and argument of periastron are kept constant at 84.6\,\kms\ and 180$^\circ$, respectively. The assumption is made that the material in the thin, eccentric ring emits equal intensity at each mean anomaly interval in its orbit. The higher velocity peak, at periastron, is less intense since the material spends less time at periastron due to the higher orbital speeds. Higher eccentricities cause the central velocity to shift away from 0 and for the $V$/$R$ ratio to lower. Interestingly, the peak-to-peak splitting remains near constant for the different eccentricities. This mirrors the variations that we observe in the emission lines which arise in the inner circumbinary ring (Figure \ref{fig:profiles_1998_2015}). Changes in the argument of periastron, $\omega$, of an eccentric disc may also lead to the variations of $v_0$ and $V$/$R$ observed in the inner ring. This is displayed in the bottom panel of Figure \ref{fig:profile_change_eccentricities}. Here, eccentricity is held constant at 0.1, and $\omega$ is varied from 0--270$^\circ$. \\

\cite{Thun2017} finds a precession period of eccentric circumbinary orbits of 
\begin{equation}
\label{eq:thun2017}
    T_{\rm prec} = \frac{4}{3} \frac{(q+1)^2}{q}\left(\frac{R}{A}\right)^{7/2} \frac{\left(1-e_{\rm d}^2\right)^2}{\left(1+\frac{3}{2}e^2\right)} P,
\end{equation}
\noindent where $T_{\rm prec}$ is the precession period, $q$ is the binary mass ratio $M_{\rm sec} / M_{\rm pr}$, $R$ is the circumbinary ring's semi-major axis, $A$ is the binary semi-major axis, $e$ is the binary eccentricity, $e_{\rm d}$ is the eccentricity of the orbiting material, and $P$ is the binary orbital period. The eccentricity of the ring about GG Car is currently unknown, but entering the binary orbital parameters gives a precession period of $\sim$\,1200 binary orbits if $e_{\rm d} \approx 0$ (which gives the longest precession timescale). The difference in time between the 1998 and 2015 observation periods corresponds to about 200 binary orbits. As a comparison, using Equation \ref{eq:thun2017}, the precession rate for the outer [O\,I] ring would be $\sim6\times 10^6$ binary orbits. It therefore follows that we would not be able to detect any significant dynamical changes in the outer ring over the FEROS observation period of $\sim$200 binary orbits. However, the FEROS observations do cover a significant portion of the expected precession period of the inner circumbinary ring, \textcolor{black}{which implies that some dynamical changes may be expected to be observed over the FEROS observations of this ring. This is as observed and displayed in Figure \ref{fig:feros_fitted_ring_parameters}, where the $v_0$ of the inner ring lines change over the FEROS observations yet the $v_0$ of the outer [O\,I] ring remains constant.} \\

We fit eccentric ring profiles to the emission lines which have been averaged for each species in each observation year, displayed in Figure \ref{fig:profiles_1998_2015}. We fit for eccentricity, $e$, argument of periastron, $\omega$, spectral resolution, $\sigma$, and velocity amplitude, alongside the parameters of continuum base level, and maximum intensity of the profile. We keep the systemic velocity constant at $-23.2$\,\kms, as was determined in Section \ref{sec:results} and corroborated by the outer [O\,I] ring.\\

\begin{figure}
  \centering
    \includegraphics[width=0.5\textwidth]{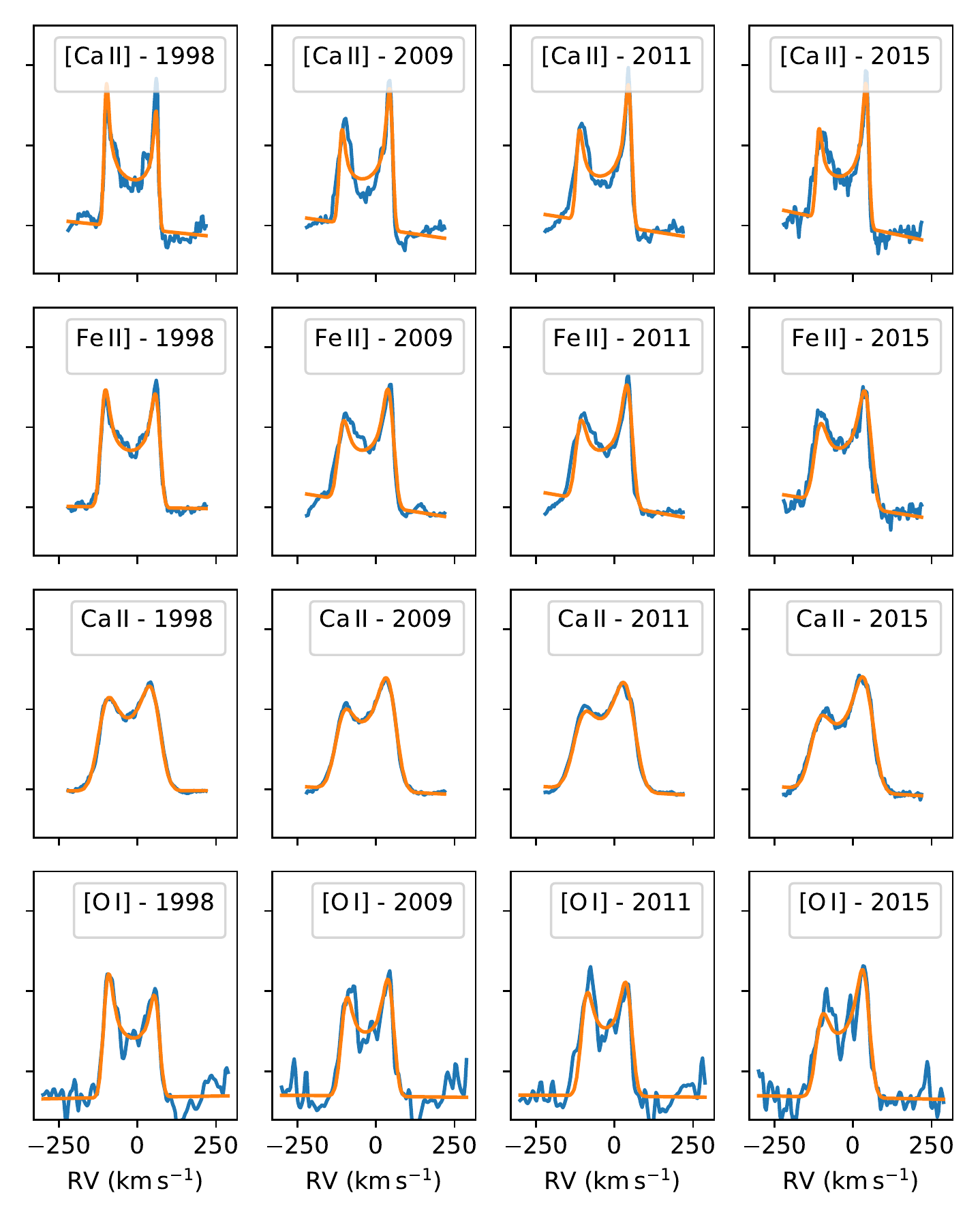}
    \caption{\textcolor{black}{Eccentric ring profile fits to the average emission line profiles for each line species in each FEROS observation year. Blue is the summed emission lines of the inner circumbinary ring in a given observation year, and orange are the fitted line profiles using the eccentric ring model. The [O\,I] line has had the profile from the outer circumbinary ring subtracted. The emission line species and observation year are given in the legend of each panel.}}
    \label{fig:all_asymmetric_ring_fits}
\end{figure}

\textcolor{black}{Figure \ref{fig:all_asymmetric_ring_fits} displays the fitted eccentric ring profiles to all studied emission lines averaged over their observation year. The line species and observation year are given in each panel. Data are shown in blue and the fitted profiles are shown in orange. All lines can be well fit using the eccentric ring model across the observation years.}\\

\begin{figure}
  \centering
    \includegraphics[width=0.5\textwidth]{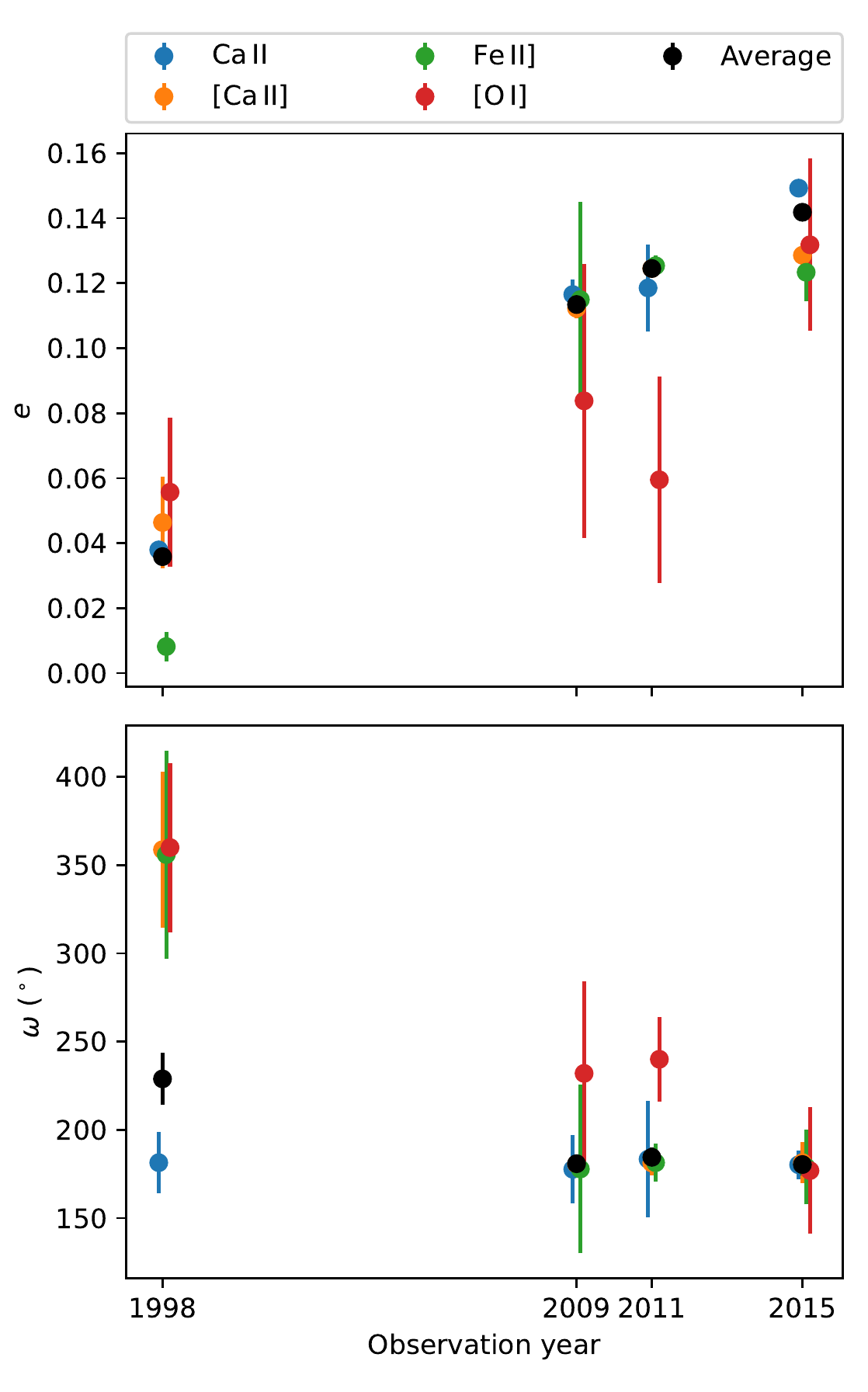}
    \caption{\textcolor{black}{Results of fitting eccentric ring profiles to the inner circumbinary ring profiles of GG Car. The top panel displays the fitted eccentricities, $e$, of the ring profiles, the middle panel displays the argument of periastron, $\omega$, and the bottom panel displays the velocity amplitude, $v_{\rm amp}$. The key in the top panel gives the line species of the data points. Black points denote the average of a fitted parameter across all line species in a given year. Data points are slightly horizontally offset for clarity.}}
    \label{fig:eccentricity_fit_results}
\end{figure}

\textcolor{black}{Figure \ref{fig:eccentricity_fit_results} displays the fitted parameters of the inner circumbinary rings for each species, along with the averaged fit parameters across all lines. For all lines we determine that the eccentricity of the inner circumbinary ring increases from $\sim$0.05 to $\sim$0.14 over the 1998 to 2015 observation period, using the eccentric ring model described. $\omega$, on the other hand, remains roughly constant at $\sim$180$^\circ$ across the years. This is due to there being some correlation between the values of $e$ and $\omega$ in the MCMC samples, since a value of $\omega \sim 180^\circ$ allows a correspondingly lower value of $e$ to recreate the asymmetrical emission line profiles observed.} \\

\textcolor{black}{Despite this slight degeneracy of $e$ and $\omega$, the fitting of the emission lines demonstrates that the changes observed in the emission lines of GG Car may be reproduced using a model of increasing eccentricity in the inner circumbinary ring. In the next section, we aim to investigate whether the time-varying potential of the binary may be induce eccentricity changes in the circumbinary ring on timescales similar to the FEROS observation window by simulating the system.}\\

\subsection{Circumbinary ring simulation}
\label{sec:phantom_simulation}
To test whether the binary may be inducing eccentricity in the inner circumbinary ring, and by extension cause the observed $v_0$ and $V$/$R$ changes of the emission lines, we simulate the binary and ring system using the smoothed particle hydrodynamic (SPH) code \phantomsph\ \citep{Price2018scpPhantom/scpAstrophysics}. \phantomsph\ has been used widely in the literature to study the dynamics of circumbinary discs.\\

We simulate the binary as sink particles with the masses and orbital parameters found in \cite{Porter2021GGPhotometry}, and given in Table \ref{tab:ggcar_parameters}. We simulate a thin, initially circular and coplanar, circumbinary ring with an initial radius of 2.8\,AU from the centre of mass of the binary. We give the ring an initial radial width of 0.01\,AU, so that it roughly matches the width of the ring estimated from the forbidden emission lines after some radial spreading (the extent of this radial spreading is discussed further in Appendix \ref{sec:phantom_hor}). Due to the very thin extent of the ring, the choices of the initial surface density and temperature power laws choices have negligible effect on the simulation. The ring has an initial mass of $5 \times 10^{-9}$\,$M_\odot$ distributed evenly between \textcolor{black}{$10^6$} SPH particles. We choose the \cite{Shakura1973BlackAppearance.} viscosity parameter, $\alpha_{\rm SS} = 0.005$; while this choice leads to a low artificial viscosity value of \textcolor{black}{$\alpha_{\rm AV} = 0.0945$}, it is acceptable for our simulation as the gas does not experience any strong shocks and it has the benefit of avoiding artificial evolution of the ring \citep{Meru2010ExploringDiscs}. In Appendix \ref{sec:phantom_alpha} we show that large changes in $\alpha_{\rm AV}$ bear little effect on the simulation output, leading us to conclude that we have chosen an appropriate artificial viscosity parameter given the aims of the simulation and its initial conditions. We use the default $\beta$ artificial viscosity value $\beta_{\rm AV} = 2$. \textcolor{black}{With these parameters, the ratio of the smoothing length to the disc scale height is \textcolor{black}{$\langle h \rangle / H $ = 0.49}}. Table \ref{tab:fiducial_simulation} displays the parameters of the simulation presented in this study. We calculate the evolution of the circumbinary ring over 4000 binary orbits.\\

\begin{table} 
\centering          
\begin{tabular}{ l l}
\hline \hline
\textbf{Binary parameters}\\
$M_{\rm pr}$ & 24\,$M_\odot$\\
$M_{\rm sec}$ & 7.2\,$M_\odot$\\
$a_{\rm bin}$ & 0.61\,AU\\
$e_{\rm bin}$ & 0.5\\
\\
\textbf{Ring parameters}\\
$R_{\rm in}$ & 2.8\,AU\\
$R_{\rm ref}$ & 2.805\,AU\\
$R_{\rm out}$ & 2.81\,AU\\
$M_{\rm disc}$ & $5 \times 10^{-9}$\,$M_\odot$\\
$H/R (R = R_{\rm ref})$ & 0.001 \\
\textcolor{black}{$\alpha_{\rm AV}$} & \textcolor{black}{0.0945} \\
$N_{\rm part}$ & \textcolor{black}{$10^6$}\\
\textcolor{black}{$\langle h \rangle / H $} & \textcolor{black}{0.49} \\
\end{tabular}
\caption{The parameters used in the circumbinary ring simulation in \phantomsph.} 
\label{tab:fiducial_simulation}      
\end{table}

\begin{figure}
  \centering
    \includegraphics[width=0.5\textwidth]{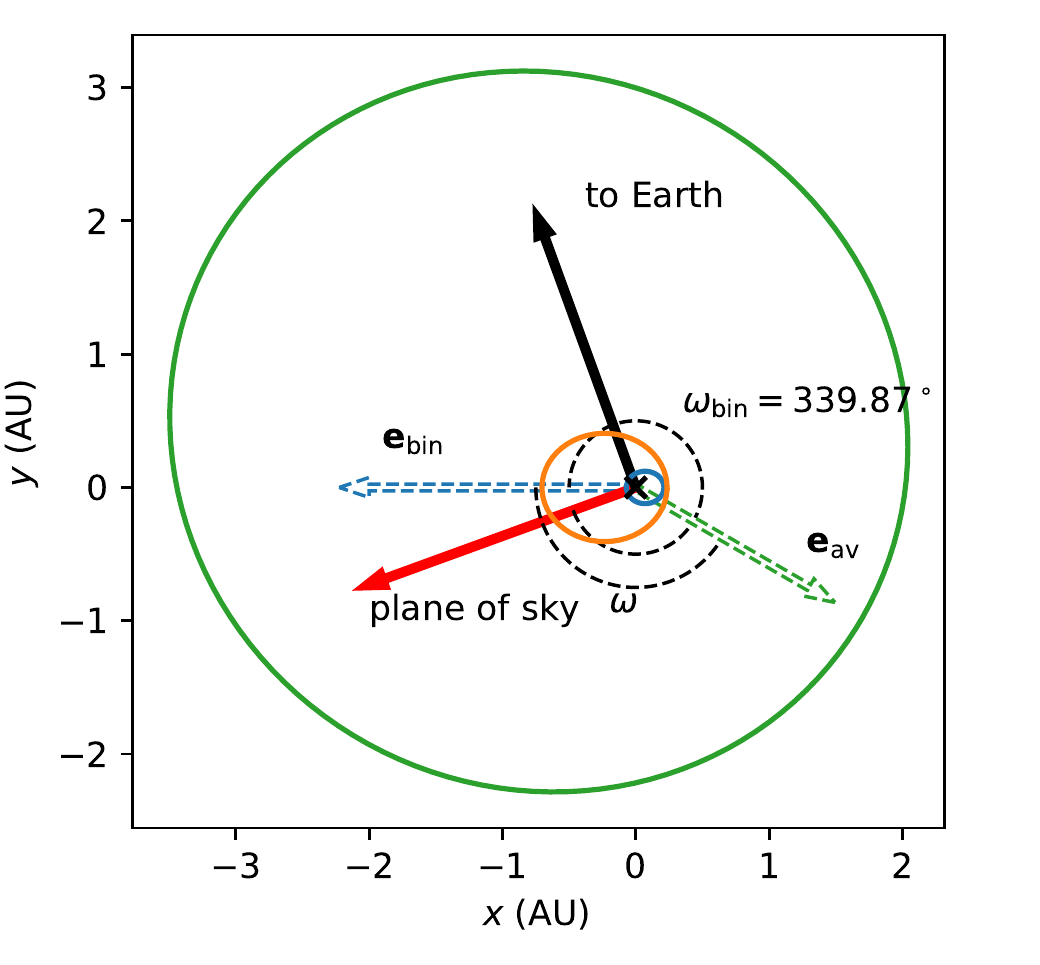}
    \caption{Schematic of the \phantomsph\ SPH simulation geometry in the $x$-$y$ geometrical plane. The blue ellipse represents the orbit of the primary star, the orange ellipse the orbit of the secondary star, and the green ellipse the orbit of the circumbinary ring at a particular instant. All three of the orbits are anti-clockwise in this schematic. The eccentricity of the ring is exaggerated in this schematic. A black cross denotes the centre of mass at $(0, 0)$. A blue arrow denotes the direction of the primary star's eccentricity vector, $\mathbf{e}_{\rm bin}$, and the green arrow denotes the direction of the circumbinary ring's eccentricity vector, $\mathbf{e}_{\rm av}$, at this particular point in the simulation. The angle from $\mathbf{e}_{\rm bin}$ to $\mathbf{e}_{\rm av}$ in the anti-clockwise direction defines $\omega$, the argument of periastron of the circumbinary ring. To project the velocities to the line-of-sight that would be seen from Earth, we utilise the argument of periastron of the primary star found in \protect\cite{Porter2021GGPhotometry}, $\omega_{\rm bin} = 339.87^\circ$; this defines the angle between $\mathbf{e}_{\rm bin}$ and the plane of the sky. The black dashed circle segments denote the angles $\omega$ and $\omega_{\rm bin}$. The plane of the sky direction is denoted by a red arrow, and the would-be line-of-sight to Earth is denoted by a black arrow.}
    \label{fig:phantom_schematic}
\end{figure}

Figure \ref{fig:phantom_schematic} displays a schematic of the simulation geometry. In this figure, the orbit of the primary and secondary stars in the $x$-$y$ plane are displayed by the blue and orange ellipses, respectively. The orbit of the circumbinary ring is displayed by the green ellipse. All orbits are in the anti-clockwise direction. We calculate the eccentricity vector of each SPH particle in the ring and take the mean of each component, giving the eccentricity vector of the ring, $\boldsymbol{e}_{\rm av}$. The eccentricity of the ring, $e = \lvert \boldsymbol{e}_{\rm av} \rvert$. We calculate the argument of periastron of the ring, $\omega$, as the angle between $\boldsymbol{e}_{\rm av}$ and the eccentricity vector of the primary star of the binary, $\mathbf{e}_{\rm bin}$, in the anti-clockwise (prograde) direction. The direction of $\mathbf{e}_{\rm bin}$ is denoted by the blue arrow and the direction of $\boldsymbol{e}_{\rm av}$ is denoted by the green arrow. \cite{Porter2021GGPhotometry} reported the argument of periastron of the primary star, $\omega_{\rm bin}  = 339.87^\circ$ to the plane of the sky; this allows us to define a direction towards Earth along which we can project velocities to calculate the radial velocities that would be observed from the Earth. The plane of the sky direction is denoted by the red arrow, and the direction to Earth is the black arrow in Figure \ref{fig:phantom_schematic}.\\

\begin{figure}
  \centering
    \includegraphics[width=0.5\textwidth]{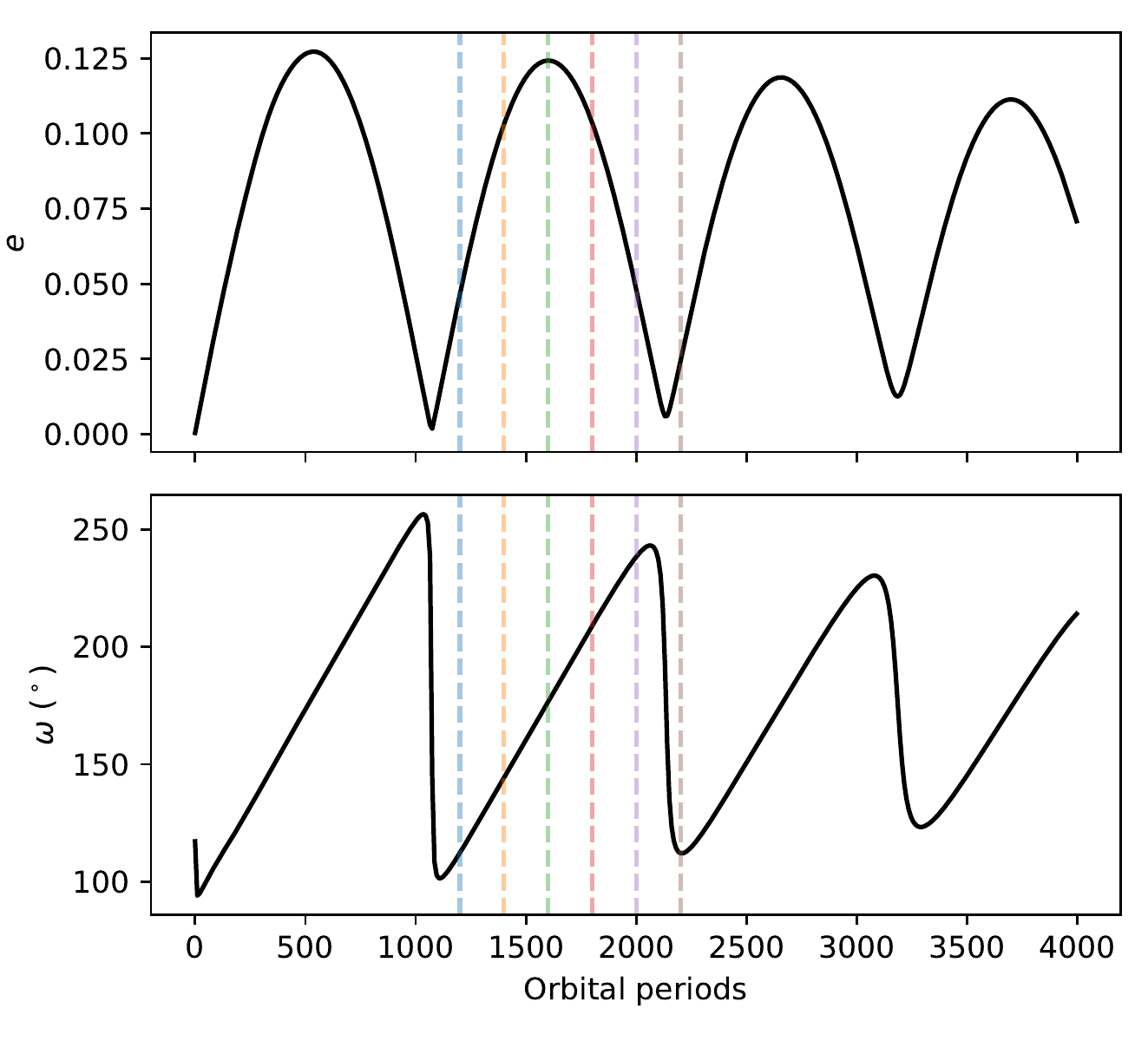}
    \caption{Eccentricity, $e$, and argument of periastron, $\omega$, of the simulated circumbinary ring against time in units of binary orbital period. $\omega$ is measured relative to the binary's argument of periastron, which is defined to be $0^\circ$. Vertical lines denote the times of the simulated emission line profiles in Figure \ref{fig:line_profiles_simulation}, with the colours of the lines matching the colours of the corresponding simulated profiles in that figure.}
    \label{fig:e_omega_simulation}
\end{figure}

Figure \ref{fig:e_omega_simulation} displays the evolution of $e$ and $\omega$ of the circumbinary ring over simulation time. The ring displays oscillations of $e$ and $\omega$, similar to those that have been observed in the previously mentioned studies of simulations of circumbinary discs. We find that the eccentricity of the ring varies between $\sim$0\,--\,0.12 with a period of $\sim$1040 binary orbits. $\omega$ varies gradually from $\sim$100\,--\,$250^\circ$, but jumps back to $100^\circ$ when $e \approx 0$. \cite{Lubow2000Interactionsdisks} predict the rate of change of eccentricity, $\dot{e}$, of a circumbinary disc about an unequal mass eccentric binary to grow as

\begin{equation}
\label{eq:lubow2000}
\dot{e} = \frac{15}{16}e_{\rm bin}\,\mu (1-\mu)(1-2\mu)\left(\frac{a_{\rm bin}}{R}\right)^3(1-e^2)^{-2}\sin \omega,
\end{equation}

\noindent where $e_{\rm bin}$ is the eccentricity of the binary, $\mu = M_{\rm sec} / (M_{\rm pr} + M_{\rm sec})$, $a_{\rm bin}$ is the semi-major axis of the binary, $R$ is the semi-major axis of the circumbinary ring's orbit.\footnote{Note that Equation \ref{eq:lubow2000} is multiplied by $-1$ compared to the Equation 2 of \cite{Lubow2000Interactionsdisks}; this is because we are measuring $\omega$ relative to the eccentricity vector of the primary, whereas \cite{Lubow2000Interactionsdisks} measure it from the eccentricity vector of the secondary. These are, by definition, 180$^\circ$ from each other, therefore $\sin(\omega+180^\circ) \rightarrow -\sin\omega$.} The relationship in Equation \ref{eq:lubow2000} between $\dot{e}$ and $\omega$ is reproduced remarkably well in the simulation, as displayed in Figure \ref{fig:e_omega_simulation}: at minimum eccentricity, $\omega$ is undefined. As the disc first becomes eccentric, $\omega \approx 100^\circ$, meaning $\sin \omega > 0$ and $\dot{e}$ is positive, leading the eccentricity to grow. As $\omega$ increases and passes 180$^\circ$, $\sin \omega$ becomes negative and $e$ begins to decrease. $\omega$ continues to increase until $e \approx 0$, at which point $\omega$ is again undefined and the process starts again.\\

\begin{figure}
  \centering
    \includegraphics[width=0.5\textwidth]{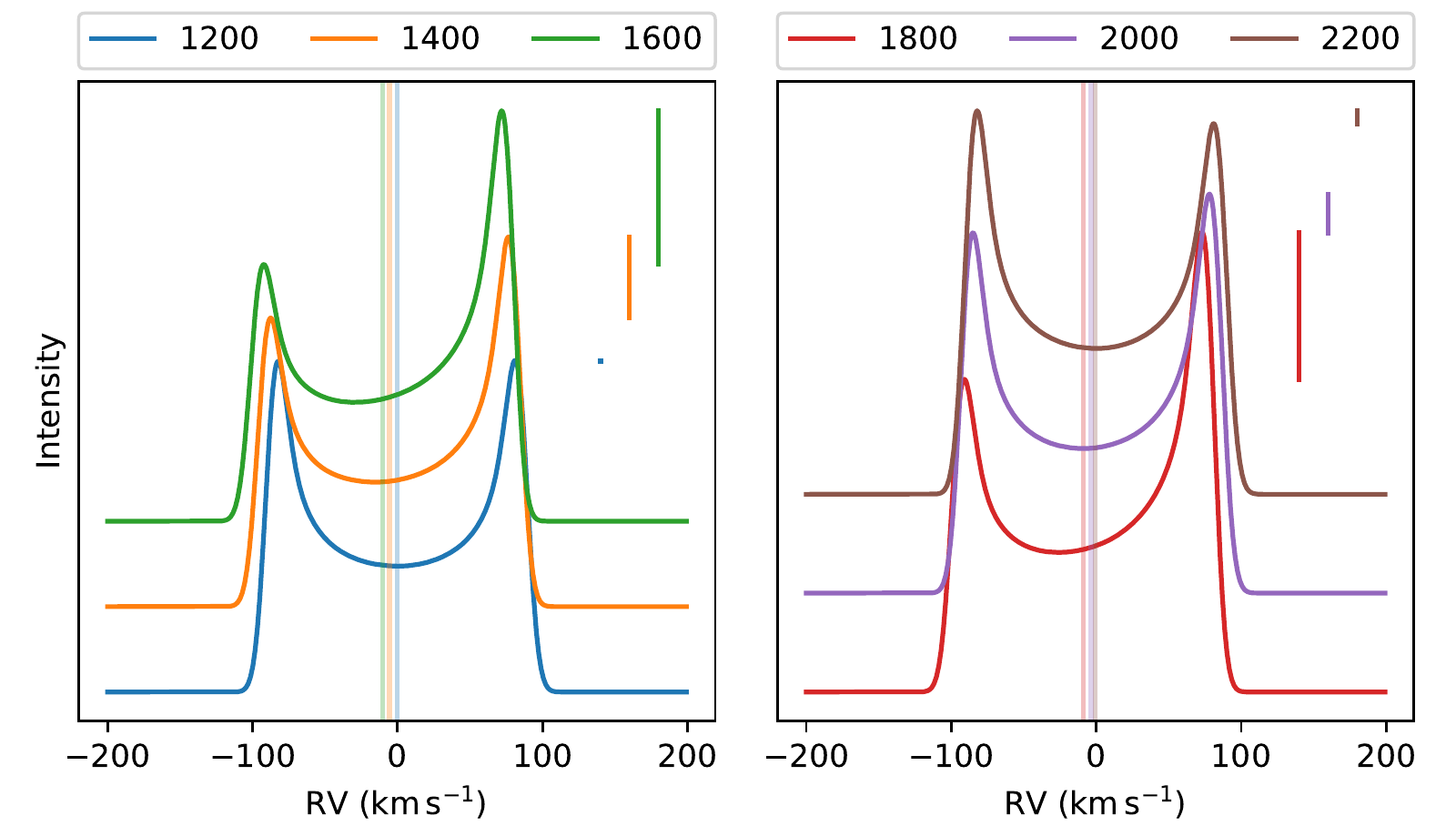}
    \caption{Simulated emission line profiles expected from the circumbinary ring at different times in the simulation. The simulation time of each profile is given in the legend, in units of binary orbital periods. These times are denoted by vertical lines in Figure \ref{fig:e_omega_simulation}, with the colours of the vertical emission lines matching its corresponding vertical line in that Figure. The velocities of the ring particles are projected such that these are the line profiles that would be observed from Earth defining the argument of periastron of the binary, $\omega = 339.87^\circ$, relative to the plane of the sky and inclination $60^\circ$. Vertical lines to the right of the profiles indicate the difference in intensity between the red peak and the blue peak of a given profile. The lines are vertically offset for clarity.}
    \label{fig:line_profiles_simulation}
\end{figure}

We calculate emission line profiles which could be expected from the circumbinary ring by projecting the velocities of the SPH particles along what would be the line of sight to Earth, demonstrated in Figure \ref{fig:phantom_schematic}, where the argument of periastron of the binary is $\omega_{\rm bin} = 339.87^\circ$ to the plane of the sky \citep{Porter2021GGPhotometry}. We multiply the projected velocities by $\sin(60^\circ)$ to account for the inclination of the disc to the line-of-sight. We histogram the radial velocities of the SPH particles and convolve with a Gaussian of width 6.25\,\kms, which is the resolution of FEROS. Figure \ref{fig:line_profiles_simulation} displays simulated line profiles corresponding to the times denoted by dashed vertical lines in Figure \ref{fig:e_omega_simulation}. The colour of the vertical line corresponds to the simulated profile in Figure \ref{fig:line_profiles_simulation}. We choose to focus on the second oscillation period to allow the simulation to reach a pseudo-steady state. The left panel shows the line profiles as the circumbinary ring becomes more eccentric, and the right panel shows the lines as the ring becomes less eccentric. The profiles are vertically offset for clarity. Vertical lines to the right of the profiles show the difference in intensity between the blueshifted peak and the redshifted peak of a given profile. The time spacing between the profiles presented is 200 binary orbits, which was chosen since this is roughly the time difference between the first and last FEROS observation. \\

Once projected along the line-of-sight that would be observed from Earth, the oscillations in eccentricity of the circumbinary ring lead to the blueshifted peak becoming less intense than the redshifted peak while the eccentricity increases, then while the eccentricity decreases the profile returns to a symmetric profile with peaks of equal intensity. The centre of the profile also shifts slightly towards negative RV while the eccentricity is increasing, returning to the centre as eccentricity decreases. The profile at 1200 binary orbital periods, in the left panel of Figure \ref{fig:line_profiles_simulation}, is symmetric despite the ring being slightly eccentric ($e \sim 0.046$); this is because $\omega \sim 110^\circ$, meaning that $\mathbf{e}_{\rm av}$ is along the line of sight to Earth, making the simulated line profile symmetric about 0\,\kms. The peak-to-peak splitting of the simulated profiles remains roughly constant throughout the oscillation cycle. The variability of the simulated emission lines in the case of increasing eccentricity closely resembles what is observed in the real circumbinary emission lines as observed by FEROS, and most clearly displayed in Figure \ref{fig:profiles_1998_2015}; over the FEROS observations, the blueshifted peak of the inner circumbinary ring profiles becomes less intense compared to the redshifted peak, and the centre of the profile shifts slightly to the blue whilst the peak-to-peak splitting remains constant. \\

\begin{figure}
  \centering
    \includegraphics[width=0.5\textwidth]{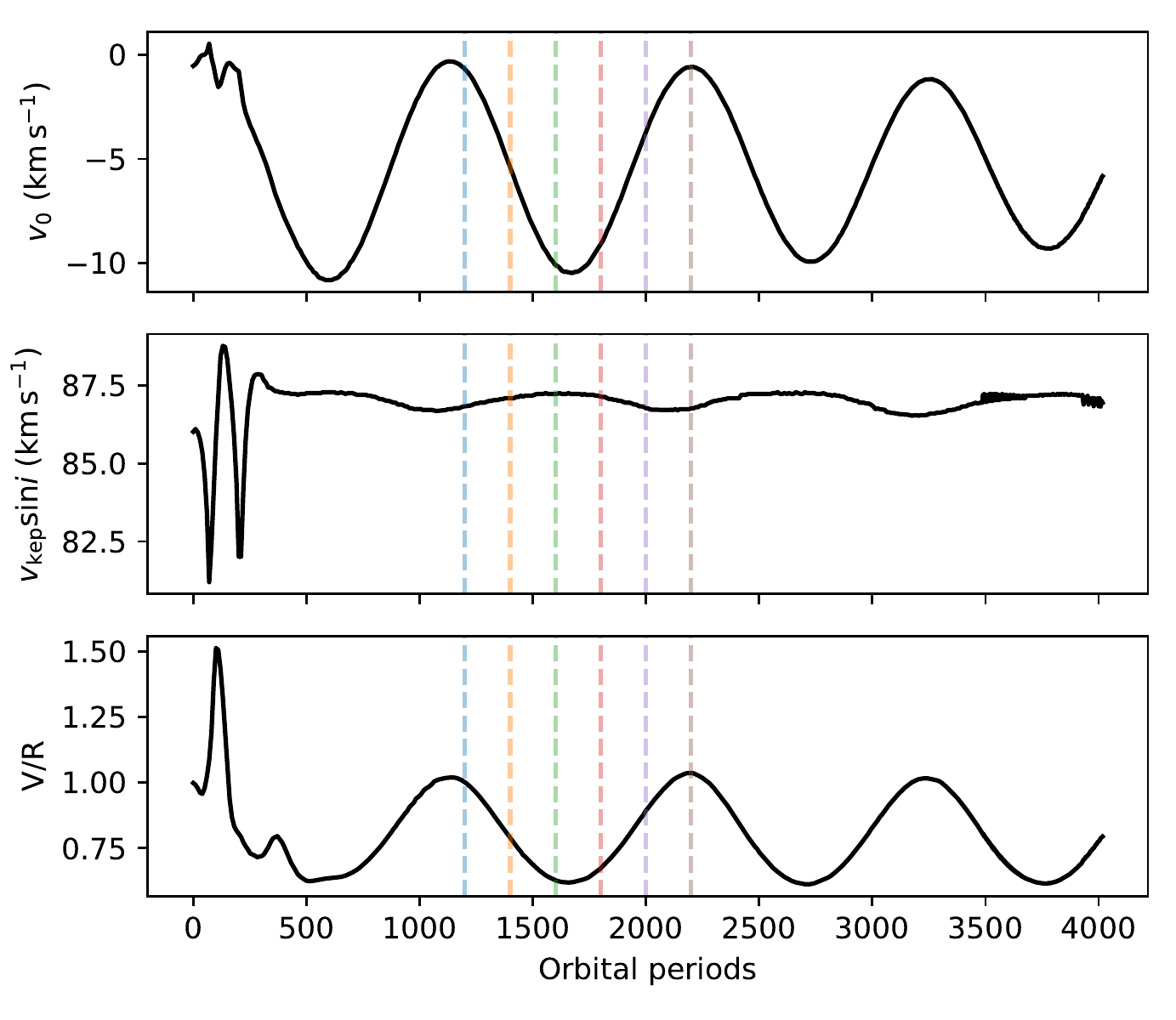}
    \caption{Results of fitting circular, symmetric ring profiles to the simulated line profiles in an identical way to the real spectra plotted against simulation time. $v_0$ is presented in the top panel, fitted $v_{\rm kep} \sin i$ is presented in the middle panel, and $V$/$R$ ratio is presented in the bottom panel. Vertical lines correspond to the same profiles as shown in Figure \ref{fig:line_profiles_simulation}, with the colours of the lines matching the colour of the simulated profiles.}
    \label{fig:v0_vkep_simulation}
\end{figure}

We have quantified the variations in $v_0$ and $v_{\rm kep}\sin i$ of the simulated emission line profiles by fitting them in an identical manner to the real line profiles as described in Section \ref{sec:model}. The fit results are displayed in Figure \ref{fig:v0_vkep_simulation}, along with the calculated $V$/$R$ ratio, in the bottom panel. Figure \ref{fig:v0_vkep_simulation} shows that it takes $\sim$500 binary orbits for the circumbinary ring to reach a pseudo-steady state. The simulated line profiles are near noiseless, which leads to more precise parameters results from the fitting than the real spectra. Figure \ref{fig:v0_vkep_simulation} shows that, using the assumption of an axisymmetric ring, the inferred systemic velocity $v_0$ varies between $\sim$\,$-10$\,--\,0\,\kms\ compared with the centre of mass of the system. $v_{\rm kep} \sin i$ displays displays some very slight variation. $V$/$R$ varies between $\sim$1 when the ring is circular, and $\sim$0.65 when the ring is most eccentric. This clearly demonstrates that the effects that we observe in the FEROS data, of decreasing $v_0$ and $V$/$R$ but seemingly constant $v_{\rm kep}\sin i$, may be caused by eccentricity oscillations in the inner circumbinary ring.\\

Although the overall magnitudes of the $v_0$ and $V$/$R$ changes, displayed in Figure \ref{fig:v0_vkep_simulation}, match that which we observe in the FEROS spectra (Figures \ref{fig:v0_1998_2015} and \ref{fig:v_o_r_1998_2015}), the FEROS data makes these changes in $\sim$200 binary orbits; for the simulated profiles, it takes $\sim$550 binary orbits for $v_0$ to change from 0\,\kms\ to $-10$\,\kms. In Appendix \ref{sec:simulation_orbital_radius} we show that, in agreement with previous work (e.g. \citealt{Lubow2000Interactionsdisks, Thun2017}), the timescale of this eccentricity oscillation is strongly dependent on the simulated circumbinary ring's initial orbital radius and that the timescale of eccentricity oscillation is much reduced with even a marginally smaller simulated initial radius. Given the large fractional error of the observed circumbinary ring's radius, this leads to a correspondingly large range of allowed oscillation timescales. In Appendix \ref{sec:simulation_orbital_radius}, we find that the allowed circumbinary ring radius, \textcolor{black}{within the uncertainties which mostly arise due to the uncertain inclination and mass of the binary}, determined from the FEROS spectra may allow for eccentricity oscillation periods in the range from $\sim$600\,--\,4000 binary orbital periods. \textcolor{black}{Given that we do not know the precise binary and ring parameters, due to their large relative uncertainties, there will therefore be a large range of allowed oscillation periods and maximum eccentricities that may be induced if we were to simulate the system across all allowed parameter space.}\\

In Appendix \ref{sec:phantom_simulation_tests} we additionally investigate the effect of varying the ring's mass, the ring's artificial viscosity, initial $H$/$R$, and simulation resolution compared to the simulation parameters listed in Table \ref{tab:fiducial_simulation}. We find that the ring's mass and artificial viscosity bears very little effect on the simulation results. A larger initial $H$/$R$ value causes the ring to spread further radially before reaching a steady state, decreasing the radius of the inner edge of the ring and therefore also decreasing the oscillation time. Varying $H$/$R$ does not affect the qualitative conclusions of this study. We therefore conclude that the SPH simulations are robust against the specific values of the various input parameters, with common behaviour which resembles the behaviour observed in GG Car existing across all plausible input ranges. \\

This shows that the variations observed in the emission line profiles originating in the inner circumbinary ring of GG Car could feasibly and plausibly be caused by the excitation of the eccentricity of the ring, and that eccentricity pumping can be expected to occur over the FEROS observation timescales given the binary's orbital solution and the radius of the inner circumbinary ring. Given this paradigm, it appears that FEROS observed GG Car as its ring was becoming more eccentric, observing just a part of an eccentricity pumping period. This scenario could be confirmed by further high-resolution observations of GG Car, which may observe the profiles becoming symmetrical and the $v_0$ of the lines returning to the systemic velocity of the system. Should a full cycle of symmetric emission line profiles, to negative $v_0$ and $V$/$R < 1$, and back to symmetry be observed within a period of $\sim$\,600\,--\,4000 binary orbital periods (50 -- 300\,years), this would give very strong evidence for eccentricity oscillations occurring in the circumbinary ring of GG Car. \\

\textcolor{black}{There are signs that for higher viscosities (Appendix \ref{sec:phantom_alpha}) or for a more radially extended ring than that simulated (Appendix \ref{sec:phantom_hor}) the eccentricity oscillations in the circumbinary ring of GG Car may be damped over a few thousand binary orbits. Therefore, the presence of eccentricity oscillations in the circumbinary ring may be a sign that the ring structures are either low viscosity or that they formed relatively recently. There is yet no definitive answer on the age or the persistence of the ring structures of \sgBeshorthand s, other than that they must have formed in the post-main sequence stages of the stars' lifetimes.}\\

\subsection{Other causes of variability}
It has previously been argued that the asymmetries of the ring profiles in \sgBeshorthand s are due to density inhomogeneities in the rings (e.g. \citealt{Maravelias2018}). Variations in the circumbinary ring profiles could therefore plausibly be caused by other means, such as dissipation of the ring leading to such inhomogeneities. Dissipation of circumbinary discs is expected due to strong illuminating radiation from the binary \citep{Alexander2012}. Other potential causes of overdensities occurring in the ring could be interaction of the ring with the wind of the primary, or reaccretion of part of the ring back onto the binary in the interim time between the two observation periods.  \\

Should the variations be due to overdensities, these could be expected to orbit about the binary, leading to the observed variations. At GG Car's inner ring's orbital radius, and with the masses of the binary, the orbital period of the ring is $310^{+140}_{-170}$\,days ($\sim$\,10 binary orbital periods), therefore variations would be expected at this period if orbiting overdensities were present. \\

Given the relative constancy of the circumbinary ring profiles between 2009--2015, it follows that the variations would be more likely to be due to eccentricity oscillations than orbiting overdensities, due to the longer timescale of the former process. As mentioned in the previous section, more epochs of high spectral resolution observations would be able to more accurately determine the variations of the circumbinary ring, and to be able to discriminate definitively between a scenario of eccentricity oscillations or orbiting overdensities. \\

\section{Conclusions}
\label{sec:conclusion}
We have presented high resolution FEROS optical spectra of the \sgBeshorthand\ binary GG Car, investigating its emission lines which arise from the circumbinary environments of the system. We find, in agreement with \cite{Maravelias2018}, that there are two concentric circumbinary rings which are manifested in GG Car's atomic emission line profiles. We have discovered that semi-forbidden Fe\,II] and permitted Ca\,II emission are formed in the inner circumbinary ring along with [O\,I], [Ca\,II], and CO emission. This is the first reported observation of permitted transition lines being observed to exhibit the ring profiles typical of forbidden emission in \sgBeshorthand s. The outer ring is only manifested in [O\,I] emission. \\

We find the inner ring has a projected orbital speed $v_{\rm kep} \sin i = 84.6\pm1.0$\,\kms, and the outer ring has $v_{\rm kep} \sin i = 27.3\pm0.6$\,\kms. Using the updated binary masses of \cite{Porter2021GGPhotometry}, and a disc inclination of $60\pm 20^\circ$, we find that the radii of the circumbinary rings are $2.8^{+0.9}_{-1.1}$\,AU and $27^{+9}_{-10}$\,AU for the inner ring and outer ring respectively. Using the circumbinary disc lines, we determine the systemic velocity of the system to be $-23.2\pm0.4$\,\kms.\\

We find no significant variability in the measured $v_{\rm kep} \sin i$ of the emission lines over $\sim$20 years of observations, indicating that the semi-major axes and inclination of the circumbinary rings are constant over this timescale. The width of the Gaussian used to convolve the disc profile for the permitted Ca\,II emission is significantly larger than for the forbidden emission lines; this implies that the emitting region of the Ca\,II emission may have a larger radial width than that of the forbidden emission.\\

We find that the inferred systemic velocity, $v_0$, from the inner circumbinary ring becomes more blueshifted over the FEROS observations, and the $V$/$R$ ratio of the double peaked profiles drops. This indicates dynamical change in the ring's orbit and a breaking of circular symmetry. \textcolor{black}{Fitting eccentric ring profiles to the emission lines indicates that these variations could be explained by eccentricity increases in the circumbinary ring.} We have simulated the inner circumbinary ring using the SPH code \phantomsph, observing that the binary can be expected to excite oscillations of the ring's eccentricity over timescales similar to the FEROS observations. We find that the variations observed in the inner circumbinary ring's emission lines observed by FEROS are consistent with the variations expected from the simulations due to these eccentricity oscillations in the case that the observations were taken whilst the eccentricity is rising, given the orientation of the system. \textcolor{black}{It is observed in the simulations that the eccentricity oscillations may become damped over many thousands of binary orbits; this may indicate that the rings are of relatively recent origin, though all that is known of the rings of \sgBeshorthand s are that they will have formed in the post-main sequence phases of the stars.} \\

Further high-resolution observations are needed to fully understand the cause of variability in the inner circumbinary ring profile of GG Car. If the ring profiles were to return to a symmetric profile with a timescale 600\,--\,4000 binary orbital periods this would give exceptional evidence to eccentricity oscillations occurring in the ring. On the other hand, cycles at $\sim$320 days may indicate that gaps in the circumbinary ring are present and their orbits about the binary cause the observed variability. We predict that the former scenario is more likely, given the relative lack of variability in the emission line profiles between 2009 and 2015. \\

It remains a mystery how these thin, circumbinary rings may be remarkably stable over the observation times given their proximity to the luminous primary star which hosts a powerful stellar wind. Further work needs to be done in order to understand the origin of these remarkable structures in \sgBeshorthand s. We have shown that, for the \sgBeshorthand s that are in binaries, the time-varying gravitational potential of the binary may lead to observable dynamical effects in their circumbinary rings.\\

\section*{Acknowledgements}
\textcolor{black}{We thank the anonymous referee for their helpful comments.} AP thanks the Science \& Technology Facilities Council for their support in the form of a DPhil scholarship, and Anna MacLaughlin and Patrick for their careful reading of the manuscript. This research has made use of NASA's Astrophysics Data System. This research has made use of the SIMBAD database, operated at CDS, Strasbourg, France.

\section*{Data availability}
FEROS spectroscopic data available from ESO at \url{http://archive.eso.org/scienceportal/home} and \url{http://dc.zah.uni-heidelberg.de/feros/q/web/form}.\\
The fits to the FEROS data underlying this article will be shared on reasonable request to the corresponding author.

\bibliography{references_new}{}
\bibliographystyle{mnras}

\appendix
\section{FEROS observations}
Table \ref{tab:feros_observations} lists the FEROS spectra of GG Car which are used in this study.\\

\begin{table} 
\centering          
\begin{tabular}{ l c}
\hline \hline
\\
Observation date & Julian Date \\
\hline
\\
1998-12-07	&	2451154.813	\\
1998-12-08	&	2451155.857	\\
1998-12-09	&	2451156.864	\\
1998-12-24	&	2451171.759	\\
1998-12-25	&	2451172.809	\\
1998-12-26	&	2451173.728	\\
1998-12-27	&	2451174.725	\\
1998-12-28	&	2451175.729 \\
2009-06-09	&	2454991.529	\\
2011-02-14	&	2455606.897	\\
2011-03-23	&	2455643.756	\\
2015-05-13	&	2457155.651	\\
2015-05-13	&	2457155.653	\\
2015-05-13	&	2457155.655	\\
2015-05-13	&	2457155.657	\\
2015-11-23	&	2457349.859	\\
2015-11-23	&	2457349.861	\\
2015-11-23	&	2457349.863	\\
2015-11-23	&	2457349.865	\\
2015-11-26	&	2457352.849	\\
2015-11-26	&	2457352.851	\\
2015-11-26	&	2457352.853	\\
2015-11-26	&	2457352.855	\\
  \hline
\end{tabular}
\caption{The FEROS observations of GG Car.} 
\label{tab:feros_observations}      
\end{table}

\section{Long term change in $v_{\rm kep} \sin i$}
\begin{figure}
  \centering
    \includegraphics[width=0.5\textwidth]{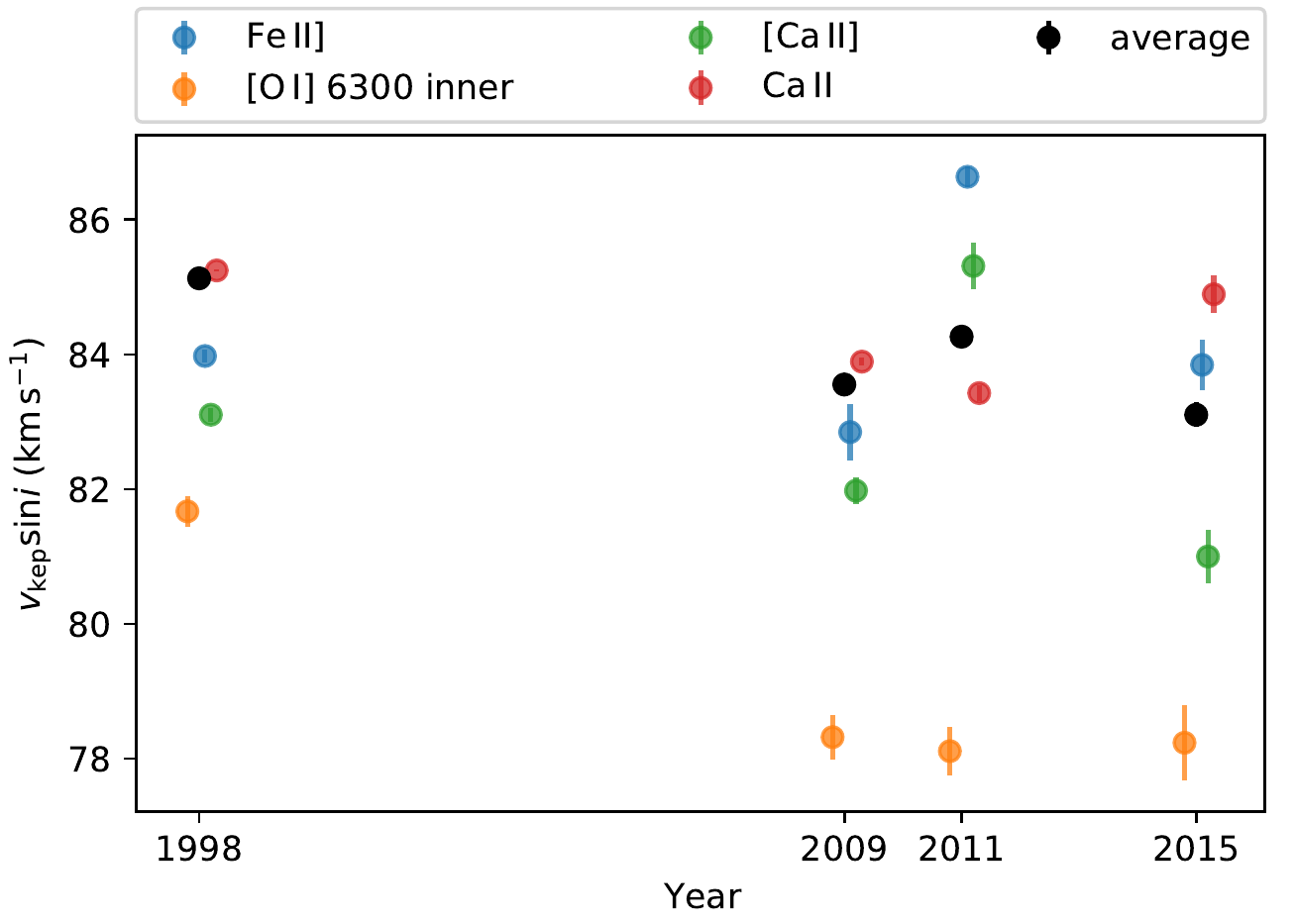}
    \caption{Same as Figure \ref{fig:v0_1998_2015}, except showing the change in $v_{\rm kep} \sin i$ in the inner ring profilesin the FEROS observations.}
    \label{fig:vkep_1998_2015}
\end{figure}

Figure \ref{fig:vkep_1998_2015} shows the fitted $v_{\rm kep} \sin i$ values for the inner circumbinary ring in GG Car in each year it was observed by FEROS. There is no significant change in the average $v_{\rm kep} \sin i$ measured, however the value for the [O\,I] line is lower in the later observations. This may be due to the inner [O\,I] ring becoming dominated by noise in the later spectra.\\

\section{Effects of varying \phantomsph\ simulation parameters}
\label{sec:phantom_simulation_tests}

We tested varying different input parameters of the inner circumbinary ring of the \phantomsph\ SPH simulations to observe their effects on the simulation output and test the robustness of our conclusions. We varied the circumbinary ring's mass, initial radius, initial $H$/$R$, artificial viscosity parameter $\alpha_{\rm AV}$, and the resolution of the simulation\\

\subsection{Resolution}
\begin{figure}
  \centering
    \includegraphics[width=0.5\textwidth]{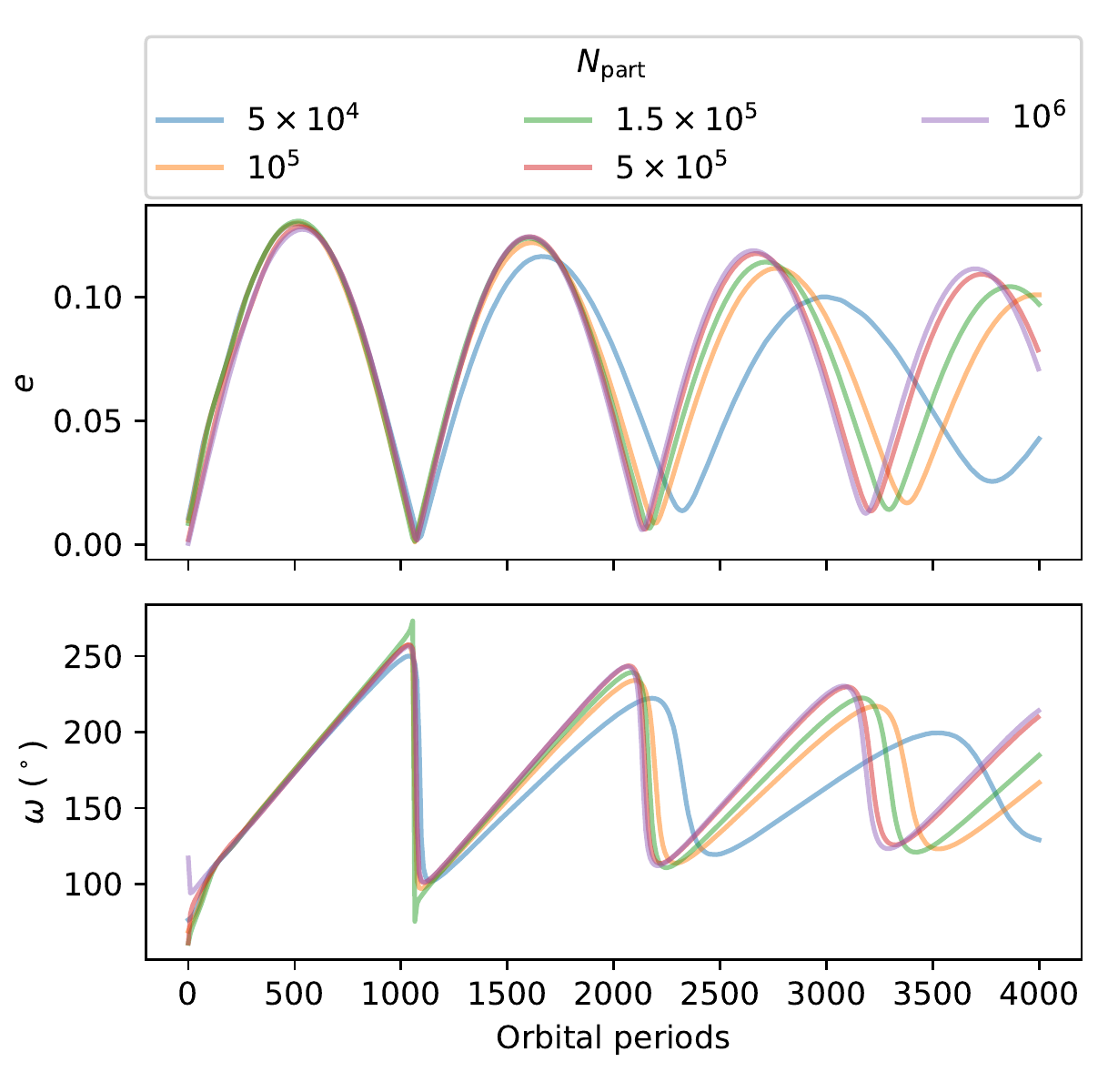}
    \caption{Effect of varying the resolution of the simulation on the oscillations of eccentricity, $e$, and argument of periastron, $\omega$, of the ring. The number of SPH particles in the simulation is shown in the legend.}
    \label{fig:e_omega_simulation_resolution_change}
\end{figure}

\textcolor{black}{Figure \ref{fig:e_omega_simulation_resolution_change} demonstrates the effect of changing the resolution of the SPH simulations on the oscillations of $e$ and $\omega$. The number of SPH particles, $N_{\rm part}$, which simulate the circumbinary ring was varied between $5\times 10^4$ and $10^6$ particles. Fewer particles allow faster computation time, but at the expense of less physically well-resolved and robust simulations. Decreasing the number of SPH particles below $10^5$ causes unacceptably divergent simulation results compared to the $10^6$ example. On the other hand, increasing the number of particles above $10^5$ does not significantly change the simulation results within the simulation time-frame but makes the calculation time much longer. Therefore, $N_{\rm part} = 10^5$ was chosen for testing the effects of varying the simulation parameters, since the simulations are accurate within the simulation time-frame and may be calculated within a reasonable calculation time.}\\

\textcolor{black}{Towards the end of the simulation time, however, there is some small divergence between the $10^6$ and $10^5$ $N_{\rm part}$ runs. This does not affect our conclusions, but if the simulations were to be run for a longer simulation time then a higher simulation resolution would be appropriate.}\\

\textcolor{black}{Varying $N_{\rm part}$ affects the simulation values of $\alpha_{\rm AV}$ and $\langle h \rangle / H$; these values for the simulations presented in Figure \ref{fig:e_omega_simulation_resolution_change} are given in Table \ref{tab:npart_alpha}.}\\

\begin{table} 
\centering          
\begin{tabular}{ l l l}
\hline \hline
\\
$N_{\rm part}$ & $\alpha_{\rm AV}$ & $\langle h \rangle / H$ \\
\hline
$10^6$ & 0.095 & 0.49\\
$5\times10^5$ & 0.075 & 0.61\\
$1.5\times10^5$ & 0.050 & 0.92\\
$10^5$ & 0.044 & 1.01 \\
$5\times10^4$ & 0.035 & 1.19\\
\end{tabular}
\caption{\textcolor{black}{The effect of varying $N_{\rm part}$ on $\alpha_{\rm AV}$ and $\langle h \rangle / H$, while keeping the other ring parameters listed in Table \ref{tab:fiducial_simulation} constant.}}
\label{tab:npart_alpha}      
\end{table}

\subsection{Ring mass}
\label{sec:phantom_disc_mass}
\begin{figure}
  \centering
    \includegraphics[width=0.5\textwidth]{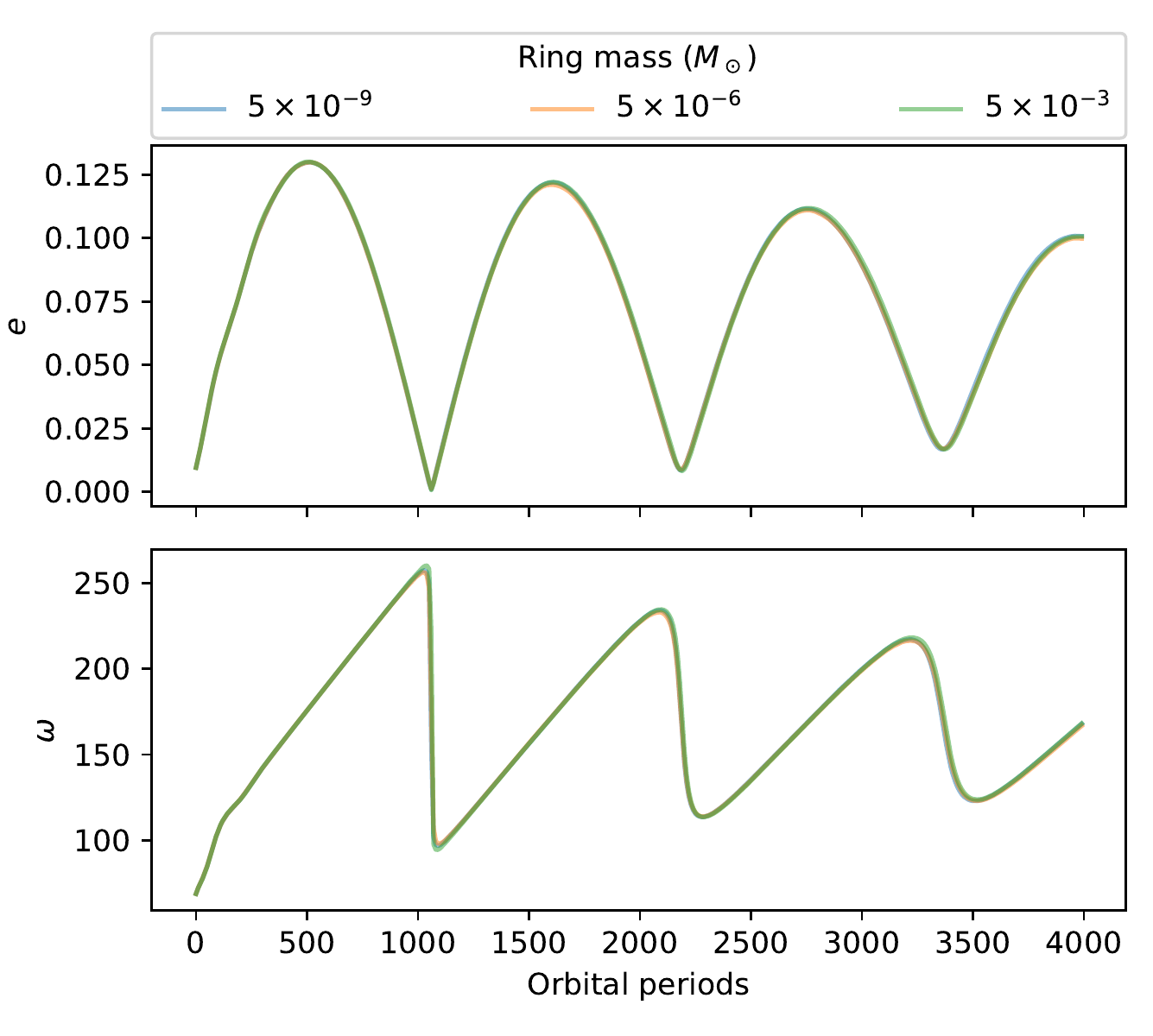}
    \caption{Effect of varying the initial mass of the inner circumbinary ring on the oscillations of eccentricity, $e$, and argument of periastron, $\omega$, of the ring.}
    \label{fig:e_omega_simulation_mass_change}
\end{figure}

Figure \ref{fig:e_omega_simulation_mass_change} displays the effect of varying the initial mass of the inner circumbinary ring in the simulations between $5\times10^{-9}$\,$M_\odot$ and $5\times10^{-3}$\,$M_\odot$. A low mass was chosen since the discs of \sgBeshorthand s are believed to be second-generation formed of material accumulated from mass-loss from the stellar surface, and would therefore have significantly lower masses than the binary components. The behaviour of the ring is near identical for each input mass, since their masses are insignificant compared to the mass of the binary components. The variation of this input parameter therefore has no significant effect on the conclusions that we draw from the SPH simulations.\\

\subsection{$H$/$R$}
\label{sec:phantom_hor}

\begin{figure}
  \centering
    \includegraphics[width=0.5\textwidth]{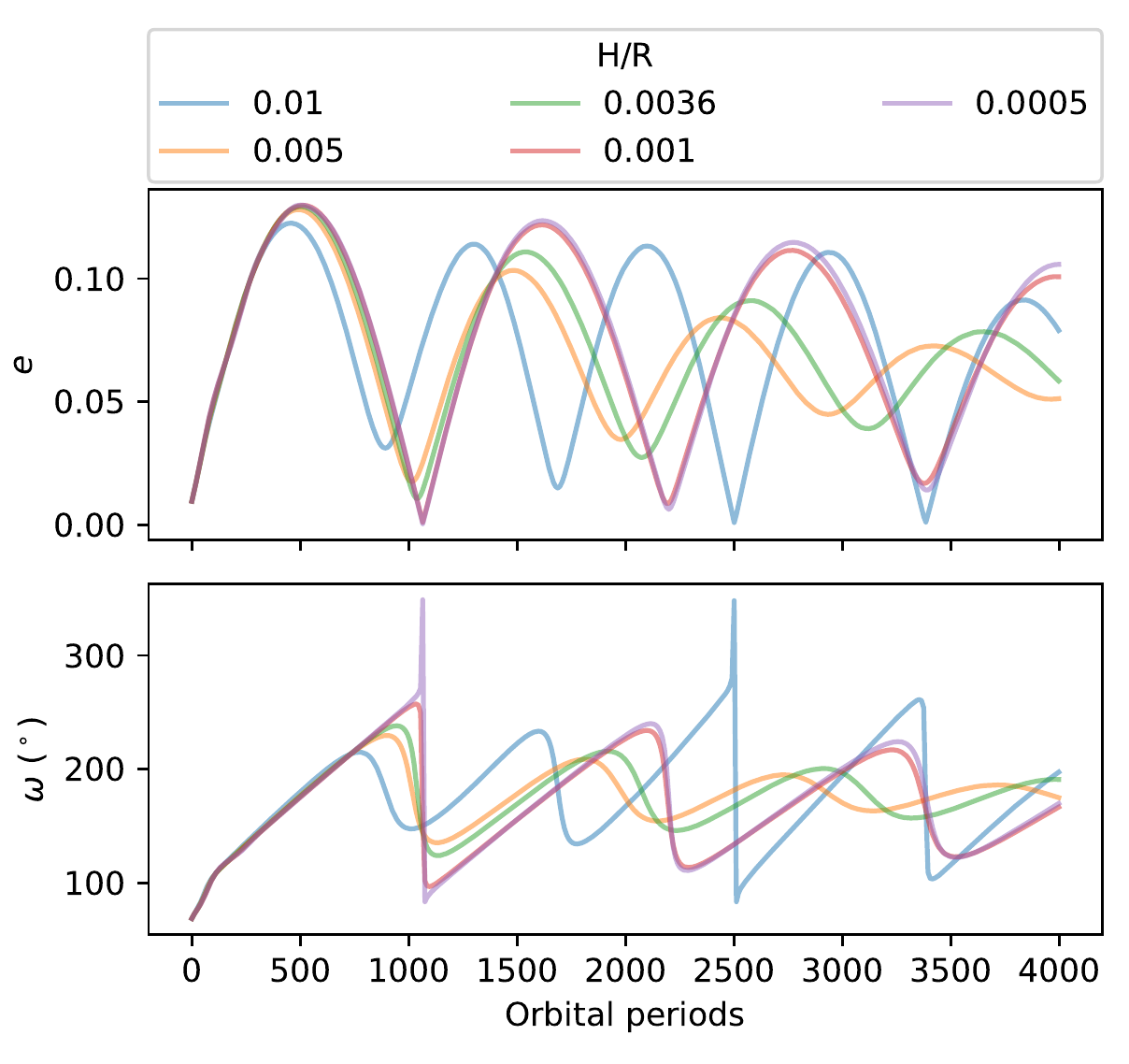}
    \caption{Effect of varying the initial $H$/$R$ of the inner circumbinary ring on the oscillations of eccentricity, $e$, and argument of periastron, $\omega$, of the ring.}
    \label{fig:e_omega_simulation_h_o_r_change}
\end{figure}

\begin{figure}
  \centering
    \includegraphics[width=0.5\textwidth]{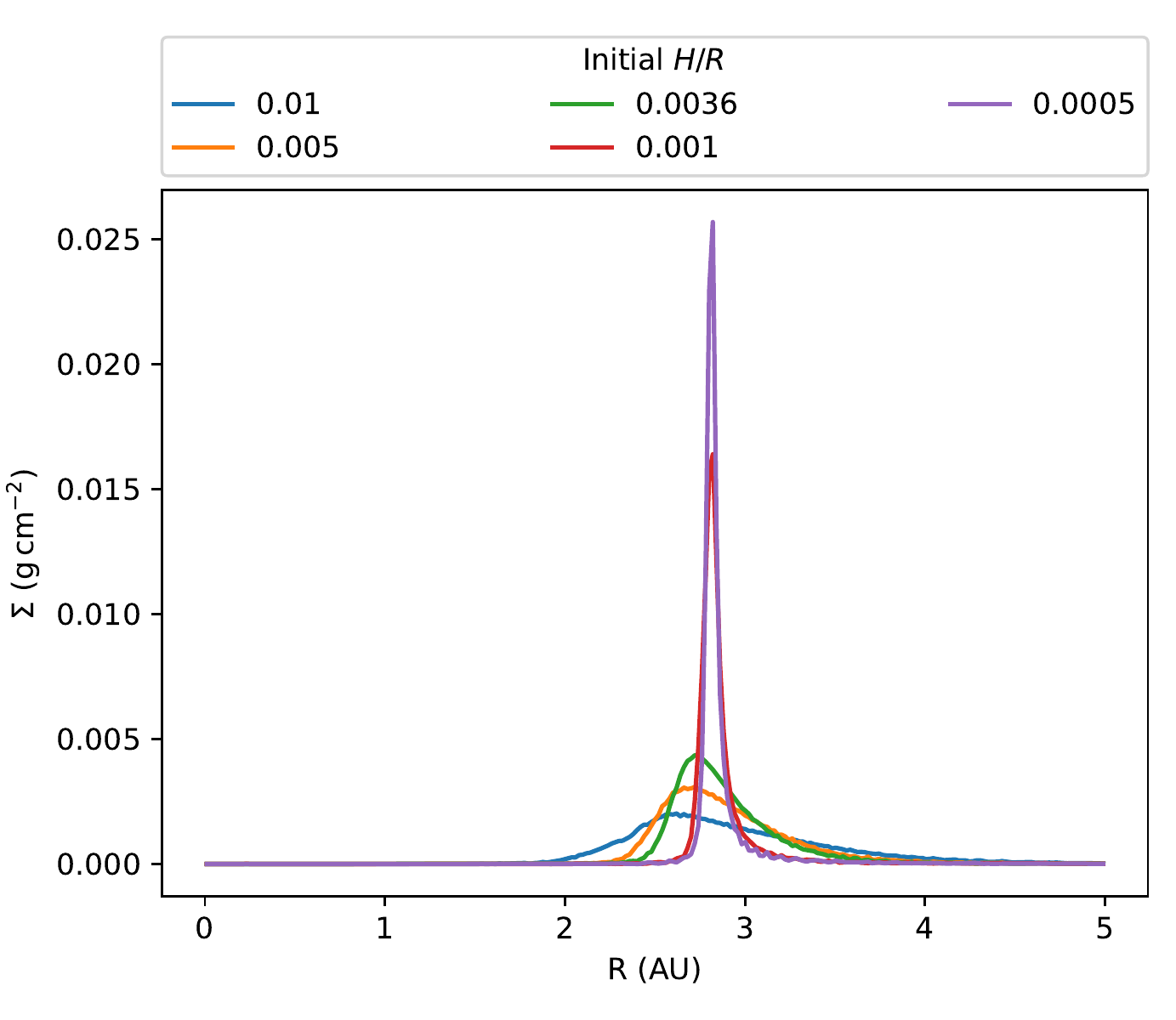}
    \caption{Surface density, $\Sigma$, of the circumbinary ring against radius, $R$, in the \phantomsph\ simulations for varying initial $H$/$R$ values at the ring's second eccentricity minimum. These $\Sigma$ distributions are calculated for the snapshots where the eccentricity of the ring is at its second minimum in Figure \ref{fig:e_omega_simulation_h_o_r_change} (i.e. at 2200 binary orbital periods for $H$/$R = 0.0005$ and 0.001; 2060 for $H$/$R = 0.0036$; 1980 for $H$/$R = 0.005$; and 1690 for $H$/$R = 0.01$). This is so that the rings are studied at similar times in their evolution after they have reached a pseudo-steady state, and when they are roughly circular.}
    \label{fig:surface_density_v_radius}
\end{figure}

Figure \ref{fig:e_omega_simulation_h_o_r_change} displays the effect of varying the initial $H$/$R$ of the inner circumbinary ring. Increasing the $H$/$R$ of the ring causes it to initially be more vertically extended. \textcolor{black}{We find that having a higher initial $H$/$R$ value causes the ring to expand radially as it reaches equilibrium, leading to a more radially extended disc distribution than a thin ring.} The distribution of the surface density of the ring, $\Sigma$, against radius, $R$, is given in Figure \ref{fig:surface_density_v_radius}; these distributions are calculated for the snapshots at the second eccentricity minimum of the circumbinary ring, such that the rings are roughly circular.\\

For a high enough value of $H$/$R$, this causes the ring to resemble more a typical circumbinary disc, rather than the thin rings observed around \sgBeshorthand s. Being more radially extended and with inner edges that are closer to the binary centre of mass, the rings with larger initial $H$/$R$ are able to couple more efficiently with the binary which causes their oscillations to have a shorter period. The oscillations in the extended ring become more rapidly damped than the narrow rings over the simulation time, matching the similar finding in \cite{Martin2018}. \\

Within the simulated times, varying the $H$/$R$ parameter has little effect on the simulated line profiles over an oscillation period, therefore our conclusions in Section \ref{sec:phantom_simulation} referring to the varying $V$/$R$ and $v_0$ of the emission lines do not strongly depend on the initial $H$/$R$ chosen. From the emission line profiles we can measure that the rings are radially thin, and therefore a smaller initial $H$/$R$ parameter is preferred. \\

\subsection{\textcolor{black}{$\alpha_{\rm AV}$}}
\label{sec:phantom_alpha}

\begin{figure}
  \centering
    \includegraphics[width=0.5\textwidth]{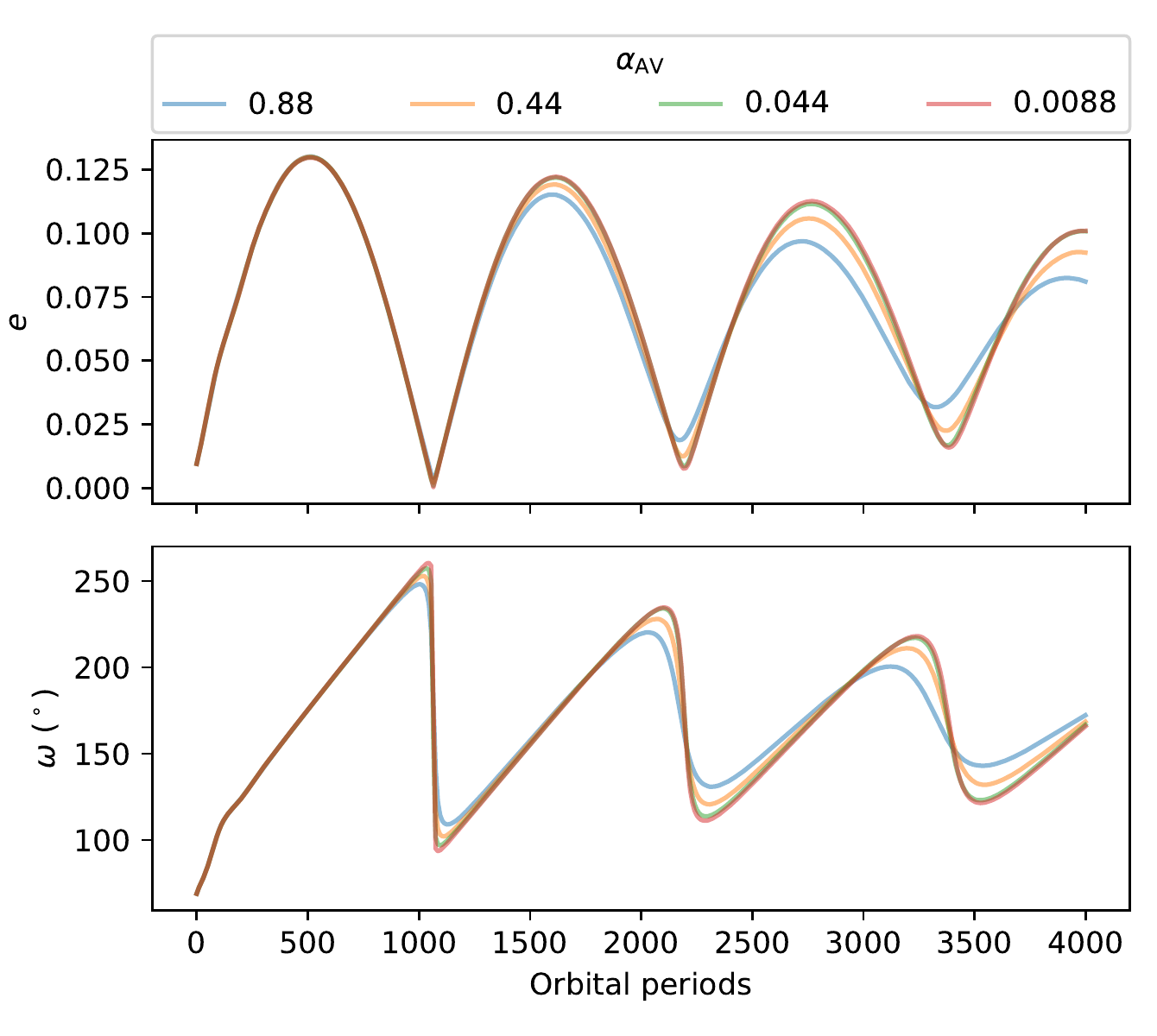}
    \caption{Effect of varying the artificial viscosity in the circumbinary ring, $\alpha_{\rm AV}$, on its oscillations of eccentricity, $e$, and argument of periastron, $\omega$.}
    \label{fig:e_omega_simulation_alpha_change}
\end{figure}

\textcolor{black}{Figure \ref{fig:e_omega_simulation_alpha_change} displays the effect of varying the artificial viscosity parameter, $\alpha_{\rm AV}$, on the eccentricity oscillations in the \phantomsph\  simulations. There is very little effect on the simulation results for large changes of $\alpha_{AV}$.} \\

\textcolor{black}{As mentioned in the main text in Section \ref{sec:phantom_simulation}, our simulation parameters lead to a low $\alpha_{\rm AV}$, though one which is acceptable given the simulation's aims and initial conditions. The low $\alpha_{\rm AV}$ value will also have the benefit of preventing artificial evolution of the ring. Increasing the artificial viscosity to a more typical value of $\alpha_{\rm AV} = 0.44$ leads to a minimal effect on the oscillations of $e$ and $\omega$ through the simulation time. If the artificial viscosity is increased to a large value of $\alpha_{\rm AV} = 0.88$ the simulation is not significantly affected, though the oscillations begin to diverge towards the end of the simulation.} \\

We interpret this as the absolute value of the viscosity, and thereby simulation artificial viscosity, having minimal effect on the simulation results, and that the simulation is robust against large changes in $\alpha_{\rm AV}$. Our conclusions on the behaviour of the circumbinary ring in GG Car are therefore valid across the investigated ranges of viscosity. \\ 

\subsection{Initial ring radius}
\label{sec:simulation_orbital_radius}
\begin{figure}
  \centering
    \includegraphics[width=0.5\textwidth]{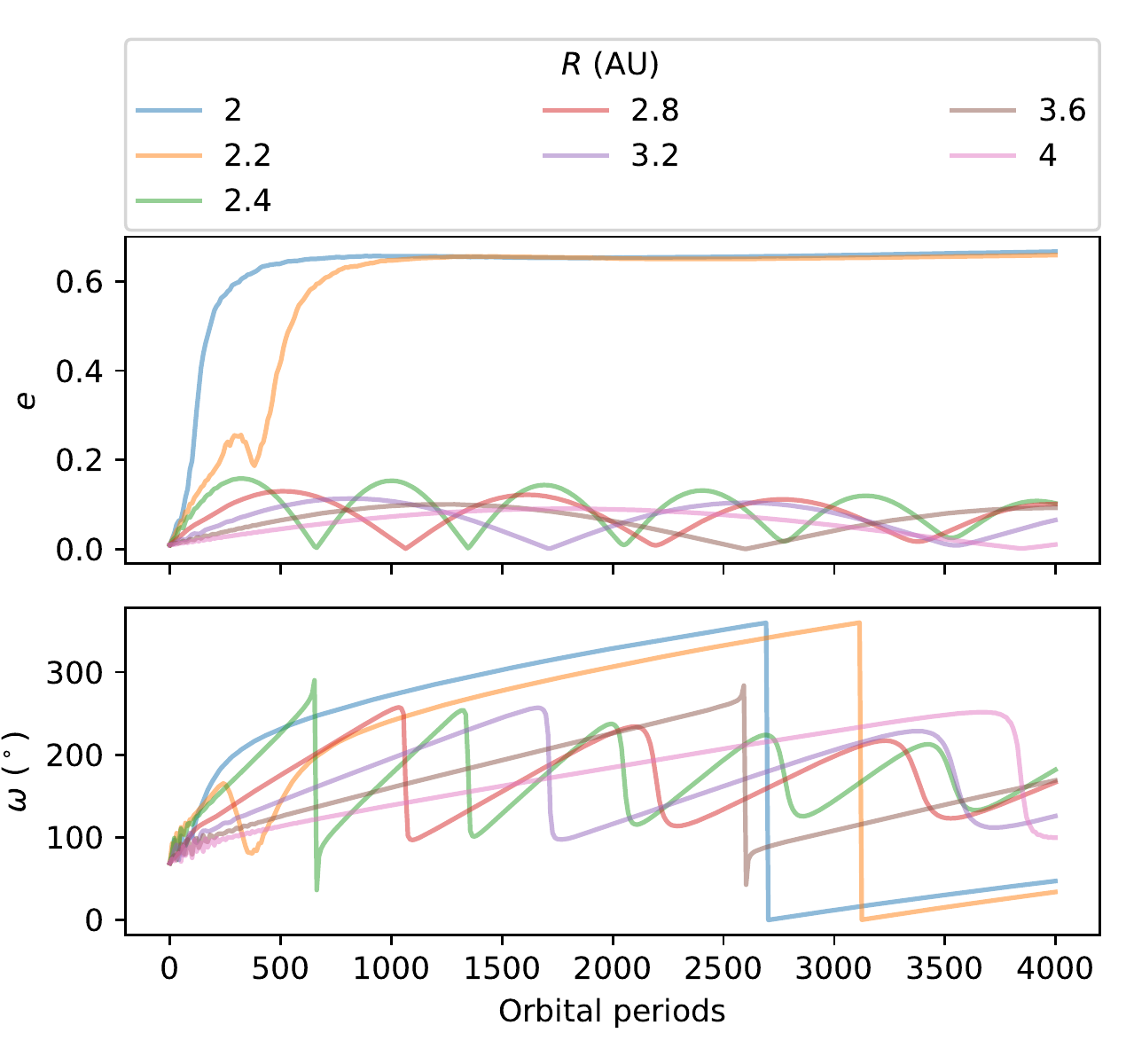}
    \caption{Effect of varying the initial orbital radius of the inner circumbinary ring, $R$, on the oscillations of eccentricity, $e$, and argument of periastron, $\omega$, of the ring in the \phantomsph\ simulation.}
    \label{fig:e_omega_simulation_orbital_radius_change}
\end{figure}

In order to check the dependence of the simulation on the initial orbital radius of the circumbinary ring, we varied the semi-major axis of the ring, $R$, between 2\,AU and 4\,AU in steps of 0.2\,AU. The 2\,--\,4\,AU range was chosen since this roughly corresponds with the limits given by the uncertainties of the radius of the inner ring. The effect of $R$ on the variations of $e$ and $\omega$ for selected models is shown in Figure \ref{fig:e_omega_simulation_orbital_radius_change}. For clarity, we do not plot the evolution of $e$ and $\omega$ for every model that we calculated.\\

As expected from previous studies (e.g. \citealt{Thun2017}), there is a strong dependence of the oscillation timescale on the initial circumbinary orbital radius. However, the closer in rings with initial radius of 2 and 2.2\,AU do not undergo oscillations in eccentricity, but are instead excited up to high eccentricities of $\sim$0.6, after which the eccentricity remains constant whilst the argument of periastron precesses.\\

\begin{figure}
  \centering
    \includegraphics[width=0.5\textwidth]{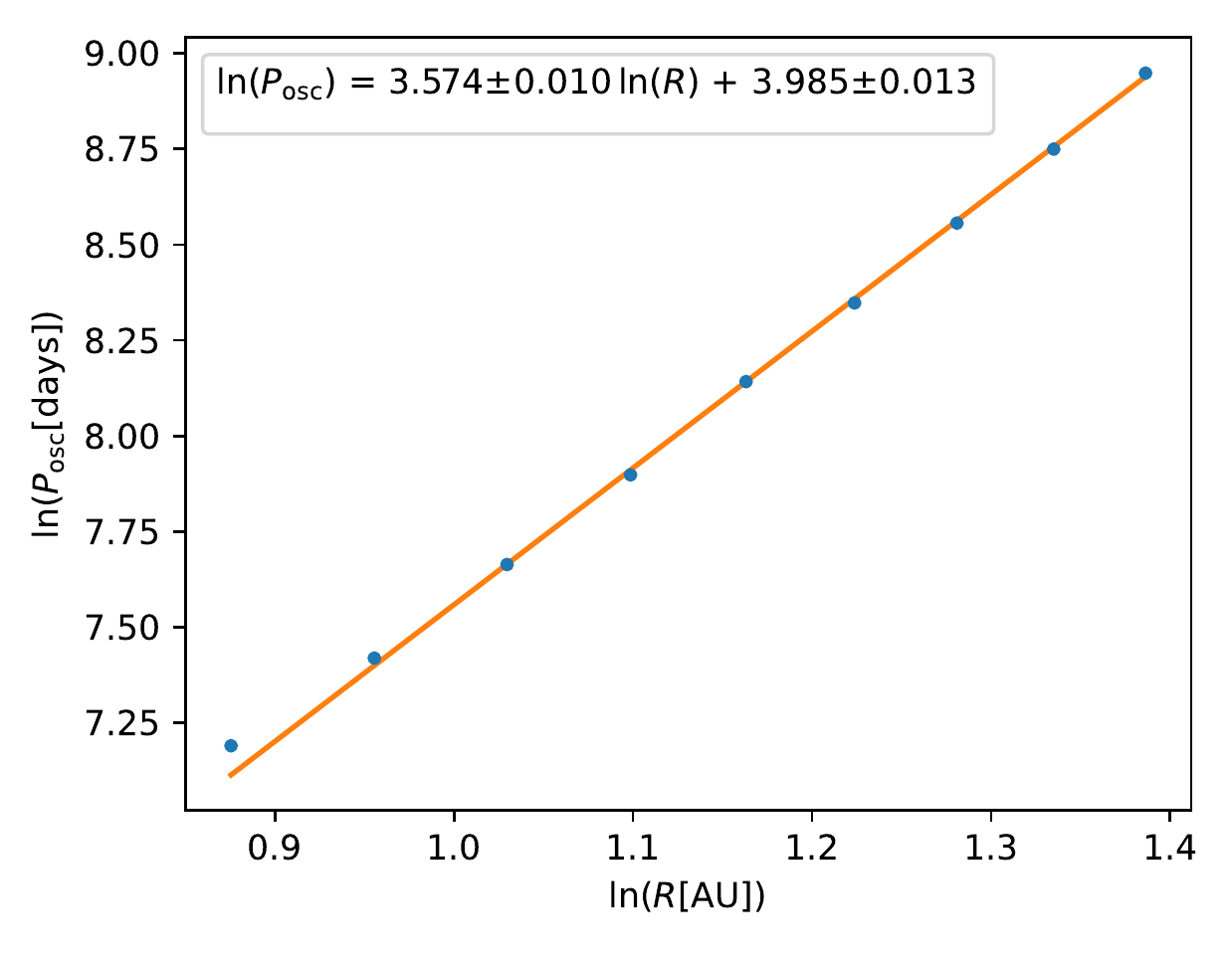}
    \caption{Relationship between the initial circumbinary ring radius, $R$, and the period of eccentricity oscillation in the ring, $P_{\rm osc}$. A linear relationship is fitted to the natural logarithms of the two parameters, and the resulting linear coefficients are given in the legend.}
    \label{fig:p_osc_v_r}
\end{figure}

For the simulations where the circumbinary ring displayed oscillations in eccentricity (i.e. simulations with $R > 2.2$\,AU), we plot the natural logarithm of the period of oscillation, $P_{\rm osc}$, against the natural logarithm of $R$ in Figure \ref{fig:p_osc_v_r}. In this instance, we define $P_{\rm osc}$ as the simulation time where the circumbinary ring reaches its first eccentricity minimum. There is a linear relationship between the two parameters in $\log$ space, and fitting a linear relationship gives a gradient of $3.574 \pm 0.01$, indicating that $P_{\rm osc} \propto R^{3.574}$. Our simulations therefore agree very well with the findings of \cite{Thun2017}, who find that the circumbinary disc precession timescale $P_{\rm prec} \propto R^{7/2}$ (Equation \ref{eq:thun2017}).

\bsp	
\label{lastpage}
\end{document}